\documentclass[11pt,fleqn]{article}
\usepackage[top=2.5cm,bottom=2.5cm,left=2.5cm,right=2.5cm]{geometry}
\usepackage[T1]{fontenc}
\usepackage[ansinew]{inputenc}
\usepackage{lmodern}
\usepackage{graphicx}
\usepackage{amsmath}
\usepackage{amsthm}
\usepackage{amsfonts}
\usepackage{indentfirst}
\usepackage{amssymb}
\usepackage{setspace}
\usepackage{textcomp}
\usepackage{float}
\usepackage{url}

\newcommand{\ba}{\begin{array}}
\newcommand{\ea}{\end{array}}

\begin{document}
\newcommand{\be}{\begin{equation}}
\newcommand{\ee}{\end{equation}}
\newcommand{\bc}{\begin{center}}
\newcommand{\ec}{\end{center}}
\newcommand{\bdm}{\begin{displaymath}}
\newcommand{\edm}{\end{displaymath}}
\newcommand{\ds}{\displaystyle}
\newcommand{\p}{\partial}
\newcommand{\INT}{\int\limits}
\newcommand{\SUM}{\sum\limits}
\newcommand{\bfm}[1]{\mbox{\boldmath $ #1 $}}
\renewcommand{\theequation}{\arabic{section}.\arabic{equation}}

\title{ \bf Mathematical modelling of variable porosity\\ coatings 
for controlled drug release
}

\author{
{\em  Sean McGinty$^{a}$, David King$^{a}$, Giuseppe Pontrelli$^{b}$} 
\vspace{5mm}\\
$^{a}$Division of Biomedical Engineering \\
 University of Glasgow, Glasgow, UK \\
E-mail: {\tt sean.mcginty@glasgow.ac.uk}\\\
\vspace{1mm}\\
$^{b}$Istituto per le Applicazioni del Calcolo - CNR \\
Via dei Taurini 19 -- 00185 Rome, Italy \\
E-mail: {\tt giuseppe.pontrelli@gmail.com}
}

\date{}

\maketitle





\begin{abstract}In this paper we investigate the extent to which variable porosity drug-eluting coatings can provide better control over drug release than coatings where the porosity is constant throughout.  In particular, we aim to establish the potential benefits of replacing a single-layer  with a two-layer coating of identical total thickness and initial drug mass.  In our study, what distinguishes the layers (other than their individual thickness and initial drug loading) is the underlying microstructure, and in particular the effective porosity and the tortuosity of the material.  We consider the effect on the drug release profile of varying the initial distribution of drug, the relative thickness of the layers and the relative resistance to diffusion offered by each layer's composition. Our results indicate that the contrast in properties of the two layers can be used as a means of better controlling the release, and that the quantity of drug delivered in the early stages can be modulated by varying the distribution of drug across the layers.  We conclude that microstructural and loading differences between multi-layer variable porosity coatings can be used to tune the properties of the coating materials to obtain the desired drug release profile for a given application.
\end{abstract}



\section{Introduction}


The topic of drug delivery is a truly multi-disciplinary research area and has been attracting the interest of engineers, mathematicians, chemists and life scientists for decades.  In particular, \textit{controlled} drug delivery has received much attention, particularly concerning the design of tablets \cite{peppas,efen,conte} and local drug delivery devices such as stents \cite{mcginty}, transdermal patches \cite{pra},  contact lenses \cite{zha} and  orthopaedic implants \cite{lyndon} (Figure 1).  Controlled release of drug from each of these vehicles can in principle be obtained by varying system design parameters.  Some of the most common include the device geometry and materials; the physico-chemical properties of the drug and; the drug loading configuration.  In the case of experimental studies, it is often \textit{demonstrated} that different drug release profiles can be obtained by either varying the experimental conditions (e.g. in-vitro versus in-vivo) or physical delivery system properties, whilst in the case of mathematical and computational modelling, it is usual for a sensitivity analysis of the underlying model parameters to be conducted, and release profiles subsequently \textit{simulated}.  Both approaches are useful and indeed can be complementary in the quest for device design optimisation. In the case of tablets, there is a body of literature concerning multi-layer systems (see e.g. \cite{peppas,efen,conte}), where the individual layers contain either different drugs or chemicals, or contrasting material properties from which the same drug or chemical is released in a bi- or multi-modal fashion.
\begin{figure}[t!]
\label{fig1}\centering\scalebox{0.7}{\includegraphics{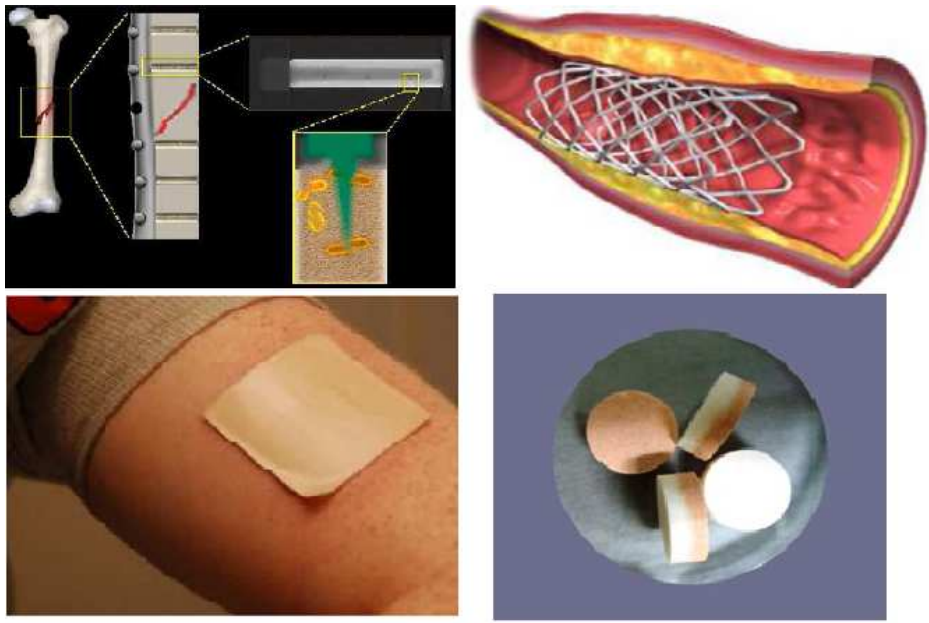}}
\caption{Examples of drug-delivery devices for different applications. From left to right: an orthopaedic implant \cite{gimeno}, a coronary stent \cite{jcr2015}, a transdermal patch \cite{patch} and  multi-layer tablets \cite{indian}. }
\end{figure}
However, the literature concerning multi-layer drug release from the aforementioned drug delivery devices is lacking somewhat, particularly in relation to mathematical modelling (see \cite{gagl}  as a rare exception). This will be the focus of the current manuscript.  


Much of the present research concerned with such drug-eluting medical devices is focussed on developing sophisticated computational  models  which  accurately  simulate  drug  release  and  the subsequent distribution in the biological environment.  The  complexity  of these models is increasing, with more and more realistic features being accounted for, including accurate 3D geometrical representations of the device and anatomical features; anisotropic and spatially-varying drug transport properties within the body and; complex features such as nonlinear binding reactions. If, on the one hand, these models are indeed necessary to accurately simulate drug transport within the device and in the biological environment, on the other hand it is clear that device manufacturers cannot intervene on the underlying biology. What they can control, however, are the properties of the device platform. Therefore, in this paper, we take a step back from the fully coupled computational models (see e.g. \cite{bozsak}) and focus instead solely on the properties of the drug-containing coating.

The  drug  is  typically  contained  within  some  durable/biodegradable  polymeric coating attached to the device platform or  embedded
within a nanoporous structure. The drug release profile depends on a number of factors including the
porosity of the coating or bulk structure; the drug loading and initial distribution; the physico-chemical
properties of the drug (e.g. molecule size, solubility, etc.) and; the release medium. A certain level of
control is required: an excessive amount of drug delivered too quickly can result in toxicity
whilst too little drug will have no therapeutic effect. However, the most desirable release profile is not always known and may in fact be
patient-specific and therapy-dependent.

Motivated by today's advances in material fabrication and by the increased capabilities of the miniaturization of structures offered by micro and nanotechnology, we propose variable porosity multi-layer coatings as an additional means of controlling the drug delivery and tailoring the release profile to the desired application. Our initial goal is to gain a better understanding of the potential benefits of replacing a single-layer with a two-layer drug-eluting coating of identical total thickness and initial drug mass.  In our study, what distinguishes the layers (other than their thickness and initial drug loading) is the underlying microstructure, and in particular the effective porosity and tortuosity of the material (Figure 2).  The structure of the paper is as follows. In Section 2 we provide the mathematical formulation of the problem and define a suitable non-dimensionalisation.  We then propose, in Section 3,  a semi-analytical solution method which makes use of separation of variables and expresses the solution as a Fourier series.  A special case which admits an analytical solution is also presented.  In the penultimate section we provide our results and investigate the sensitivity of the release profile to variations in the model parameters.  Finally, in Section 5, we provide the conclusions of our study.
\begin{figure}[t!]
\label{fig2} \centering\scalebox{0.4}{\includegraphics{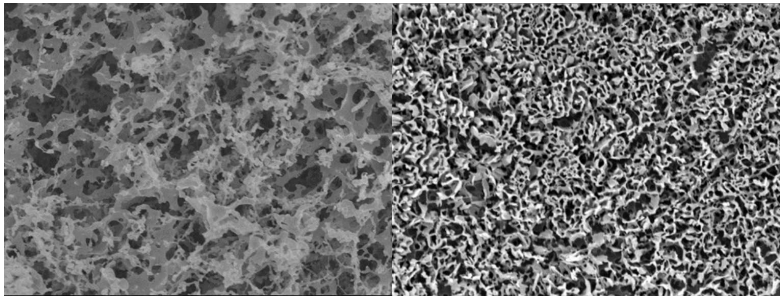}}
\caption{Example of two adjacent polymer coatings with different microstructural properties. These were prepared from different concentrations of polymer solutions (0.6\% left and 0.8\% right) \cite{teng}.}
\end{figure}

\section{Mathematical formulation}
\setcounter{equation}{0}

A drug delivery device typically includes a polymeric matrix coating containing drug which is
 in contact with some release medium.  The particular geometry of the device varies between applications, but the drug-eluting coating can usually reasonably be idealised as a slab (\textit{layer}) of some thickness $L$. In Figure 2 we display an example of the situation we wish to model in the present work: two adjacent coating layers with different microstructural properties.  Since the total thickness of drug-eluting coatings is typically small relative to the lateral coating dimensions, and the net drug transport is along a single direction, we restrict our attention to a one-dimensional model (Figure 3).  We consider layers 1 and 2 to  have thickness $L_1$ and $L_2$, respectively,  with $L=L_1+L_2$ the total coating thickness, which we keep fixed in the following.
\begin{figure}[h]
\centering\scalebox{0.3}{\includegraphics{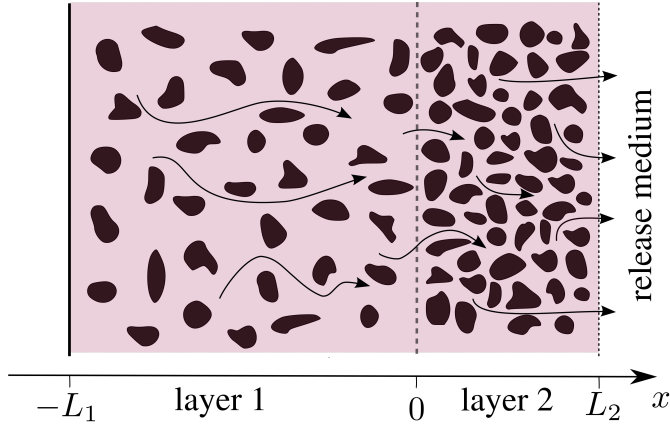}}
\caption{Schematic showing a simplified one-dimensional configuration of
drug release from a medical device coating into a release medium. Two porous layers of different
thickness and structure are faced.
Due to an initial difference of drug concentrations between
the two layers and the release medium, a mass flux is established  to the right and
  drug diffuses through the coating and into the release medium.
 Figure not to scale.}
\label{fig3}
\end{figure}
We represent each layer of the porous coating as a homogeneous material and define some representative
elementary volume (r.e.v.) of  size larger than the pore
scale, but smaller than the typical length scale of the phenomenon.  Within the r.e.v., we have solid and void parts.  We choose to define all concentrations as intrinsically averaged variables, that is, averaged with respect to the void volume, rather than the total r.e.v.

Let $c_1$ and $c_2$ then denote the intrinsic concentrations of drug in layer $1$ of constant porosity $\phi_1$ and layer $2$ of constant porosity $\phi_2$, respectively.  We further define $\phi_i^e, \; i=1,2$ ($0<\phi_i^e\leq \phi_i$) as the effective transport-through porosities, which may be smaller than the overall porosity of each layer if, for example, there are small inaccessible pores or dead-end pores \cite{grat}.  Additionally, we directly account for the fact that the molecules may have to travel through an increased path length due to the circuitous nature of the pores by introducing a tortuosity parameter $\tau_i, \; i=1,2$.  Assuming that the coating is rapidly wetted and that the drugs are readily soluble, it can be shown  that drug transport satisfies the following diffusion equations
\begin{eqnarray}
\label{1} \phi_1\frac{\partial c_1}{\partial t}&=&D_1^e \frac{\partial^2 c_1}{\partial x^2}, \;\;\quad -L_1<x<0, \quad t>0, \\
\label{2} \phi_2 \frac{\partial c_2}{\partial t}&=&D_2^e \frac{\partial^2 c_2}{\partial x^2}, \quad\quad 0<x<L_2,\quad t>0,
\end{eqnarray}
where $D_1^e=\ds\frac{\phi_1^e D^w}{ \tau_1}$ and $D_2^e=\ds\frac{\phi_2^e D^w}{\tau_2}$ are the effective diffusion coefficients in each layer and $D^w$ is the corresponding free diffusion of drug in water \cite{cussler} . We emphasize that $D^w$ is independent of the microstructure and that we consider only the case of the same drug in each layer. In this work we envisage medical implant coatings which  release drug through fluid-filled pores \textit{only}.  As a consequence, we do not consider diffusion in the solid phase, which can be several orders of magnitude slower than in the liquid phase.

For the sake of generality, we impose  a mixed-type condition at
both ends:
\begin{eqnarray}
\label{31}-D_1^e \frac{\partial c_1}{\partial x}&=& K_1 c_1, \;\;\quad\qquad\qquad x=-L_1, \quad t>0,\\
\label{61}-D_2^e \frac{\partial c_2}{\partial x}&=& K_2 c_2, \;\;\quad\qquad\qquad x=L_2, \quad t>0,
\end{eqnarray}
where we may, in principle, choose $K_1$ and $K_2$ to match experimentally measured flux.
The above boundary conditions allow us to explore the two extreme cases of zero flux and infinite sink conditions (see Section 4).   If, for example, the coating is attached to an impermeable device (e.g. a stent) and drug release is measured under infinite sink conditions, we can let $K_1=0$ and $K_2\rightarrow \infty$.

At the interface between the two layers we impose continuity of flux. To keep the problem general, this flux accounts
 for a possible drug partitioning or a non-perfect contact, modelled through a mass transfer
coefficient $P (m/s)$:
\begin{eqnarray}
\label{4} -D_1^e\frac{\partial c_1}{\partial x}& =& P(c_1-c_2), \;\;\quad\qquad\qquad x=0, \quad t>0,\\
\label{5}-D_1^e\frac{\partial c_1}{\partial x}&=&-D_2^e\frac{\partial c_2}{\partial x}, \;\;\;\;\;\quad\qquad\qquad x=0, \quad t>0,
\end{eqnarray}
We assume that initially the drug is loaded at uniform concentrations $c_1^0$ and $c_2^0$ in layers 1 and 2, respectively:
\begin{align}
&c_1=c_1^0, \quad -L_1\leq x\leq 0, \quad\qquad t=0,  \nonumber  \\
&c_2=c_2^0, \quad  0< x\leq L_2, \quad\qquad  t=0. \label{ert}
\end{align}
The case of a single layer can easily be recovered, as will be demonstrated in Section 4.

\subsection{Non-dimensionalisation}
\noindent We now proceed to non-dimensionalise equations (\ref{1})-(\ref{ert}).
We choose
\bdm
x'=x/L, \qquad t'=D_1^e t/\phi_1L^2, \qquad c_1'=c_1/c_1^0, \qquad c_2'=c_2/c_1^0.
\edm
The non-dimensionalised equations (after dropping primes) are then:
\begin{eqnarray}
\label{1'} \frac{\partial c_1}{\partial t}&=& \frac{\partial^2 c_1}{\partial x^2}, \;\quad\quad\quad\quad -\delta<x<0, \quad t>0, \\
\label{2'} \frac{\partial c_2}{\partial t}&=&\ds{\chi \over \phi} \frac{\partial^2 c_2}{\partial x^2}, \quad\quad\quad\quad 0<x<1-\delta,\quad t>0, \\
\label{3'}-\frac{\partial c_1}{\partial x}&=&\Gamma_1 c_1, \quad \qquad\qquad x=-\delta_, \quad t>0,\\
\label{4'} - \frac{\partial c_1}{\partial x}&=& \Pi (c_1-c_2), \;\;\qquad x=0, \quad t>0,\\
\label{5'} \frac{\partial c_1}{\partial x}&=&\chi  \frac{\partial c_2}{\partial x}, \;\;\quad\quad \qquad x=0, \quad t>0,\\
\label{6'} -\chi  \frac{\partial c_2}{\partial x}&=& \Gamma_2 c_2, \;\;\;\qquad\qquad x=1-\delta,\; t>0,
\end{eqnarray}
\begin{align}
&c_1=1, \quad -\delta \leq x \leq 0, \qquad t=0.\nonumber \\
&c_2=C^0, \quad 0< x \leq 1-\delta,   \qquad t=0.  \label{7'}
\end{align}
where
\begin{align}
&\delta=\frac{L_1}{L}, \qquad  \chi= {D_2^e \over D_1^e},\qquad  C^0={c_2^0 \over c_1^0},   \qquad  \phi=\frac{\phi_2}{\phi_1},
&\Pi={ P L  \over D_1^e}, \qquad \Gamma_1={K_1 L \over D_1^e},   \qquad
\Gamma_2={ K_2 L  \over D_1^e}. \nonumber
\end{align}
We note that the non-dimensional parameter $\chi$ contains all the important microstructural parameters which influence drug release. \\

\section{Solution procedure}
\subsection{Solution by Separation of Variables}
\setcounter{equation}{0}
The model given by (\ref{1'})-(\ref{7'}) is amenable to solution by separation of variables, an approach we have adopted in previous work considering two-layer and multi-layer problems \cite{pontrelli3, pon2}. We let
\be
c_1(x,t)=X_1(x) G_1(t),   \qquad\qquad c_2(x,t)=X_2(x) G_2(t). \label{eq1}
\ee
Equations (\ref{1'})--(\ref{2'}) give rise to the ordinary differential equations (ODEs):
\be
 {G_1' \over G_1} =-\lambda_1^2,
\qquad\qquad
{\phi \over \chi} {G_2' \over G_2} =-\lambda_2^2,  \label{abc1}
\ee
which yield the solution:
\be
G_1(t)= \exp({-\lambda_1^2 t}),  \qquad\qquad\qquad\qquad  G_2(t)=
\exp\left({-{\chi \over \phi}\lambda_2^2 t}\right),
\label{sec3}
\ee
and the Sturm-Liouville eigenvalue system:
\begin{eqnarray}
X_1''&=& -\lambda_1^2 X_1,  \qquad   \, \, -\delta<x<0,  \label{sl1}\\
-X_1'&=&\Gamma_1 X_1,  \qquad   \, \, \;\;  x=-\delta, \label{sl2} \\
X'_1&=& \chi X'_2, \qquad   \,\,\;\; \;\; \label{sl3} x=0,
\end{eqnarray}
\begin{eqnarray}
X_2''&=& -\lambda_2^2 X_2,   \qquad   \, \,\;\;\; 0<x<1-\delta, \label{sl4} \\
 -\chi X_2'&=& \Gamma_2 X_2,   \qquad   \, \,  \, \;\;\;\;\;x=1-\delta,  \label{sl5} \\
- X_1'  &=& \Pi(X_1 - X_2), \quad   \, \, x=0,  \label{sl6}
\end{eqnarray}
obtained by setting $G_1=G_2$, which implies
\be
  \lambda_1 = \sqrt{\chi \over \phi } \, \lambda_2.   \label{pio}
\ee

\noindent The general solution of the ODEs (\ref{sl1}) and (\ref{sl4}) is:
\be
 X_1(x)=
a_1\cos(\lambda_1 x) + b_1 \sin(\lambda_1 x),  \qquad\qquad
 X_2(x)= a_2\cos(\lambda_2 x) + b_2 \sin(\lambda_2 x),  \label{sec31}
\ee
where the eigenvalues $\lambda_i$ ($i=1,2$) and the unknown coefficients $a_i$ and
$b_i$ may be computed by
imposing the  boundary and interface conditions as follows.
From equations (\ref{sl2}) and (\ref{sl5}), we have:
\begin{eqnarray}
a_1 ( \lambda_1 \sin(\lambda_1 \delta) + \Gamma_1 \cos(\lambda_1 \delta))  + b_1 ( \lambda_1 \cos (\lambda_1 \delta )  - \Gamma_1 \sin(\lambda_1 \delta)) &=&0, \nonumber \\    \label{sec41}  \\
 a_2 [-\chi \lambda_2  \sin(\lambda_2 (1-\delta)) + \Gamma_2 \cos(\lambda_2 (1-\delta))]    +
 b_2 [ \chi
\lambda_2 \cos(\lambda_2 (1-\delta)) + \Gamma_2 \sin(\lambda_2 (1-\delta))  ]&=&0. \nonumber \\    \label{sec42}
\end{eqnarray}
From the interface conditions (\ref{sl3}) and (\ref{sl6}), it follows:
\begin{eqnarray}
 b_1 \:\lambda_1  &=& \chi \: \lambda_2 \: b_2,  \label{sec7} \\
 -b_1 \lambda_1 &=& \Pi (a_1 - \: a_2) \label{sec62}.
\end{eqnarray}

Equations (\ref{sec41})--(\ref{sec62}) form a system of four
homogeneous linear algebraic equations in the four unknowns $a_1, b_1, a_2$  and $b_2$ .
To obtain a solution different from the {\em trivial} one $(0,0,0,0)$, it
is necessary that the determinant of the coefficient matrix
associated with the above system is equal to zero, that is:
 \begin{align}
&\varphi(\lambda_1, \lambda_2)= \bigg( \lambda_1 \sin(\lambda_1 \delta) + \Gamma_1 \cos (\lambda_1 \delta) \bigg) \,
\bigg[  \Pi \bigg(\chi \lambda_2 \cos(\lambda_2 (1-\delta)) + \nonumber \\
& \Gamma_2  \sin(\lambda_2 (1-\delta))\bigg) +
  \chi \lambda_2 \bigg(-\chi \lambda_2 \sin(\lambda_2 (1-\delta)) + \Gamma_2  \cos(\lambda_2 (1-\delta))\bigg)  \bigg] -   \nonumber \\
& \Pi \sqrt{\chi \phi} \bigg( \lambda_1 \cos(\lambda_1 \delta) - \Gamma_1 \sin (\lambda_1 \delta)\bigg)
\bigg( - \chi \lambda_2 \sin(\lambda_2 (1-\delta)) + \Gamma_2 \cos (\lambda_2 (1-\delta))\bigg)    =0  \label{se11}
\end{align}
By replacing $\lambda_1$ with $\lambda_2$ through the relation (\ref{pio}),
if the above transcendental equation (eigen condition) in $\lambda_2$  is satisfied,
the coefficients may be taken as:
\begin{eqnarray}
a_2&=&    {\chi \lambda_2 + \Gamma_2 \tan ( \lambda_2 (1-\delta)) \over
- \Gamma_2 + \chi \lambda_2 \tan(\lambda_2 (1-\delta))  }    b_2, \label{se8}\\
a_1&=& a_2- {\chi \over \Pi} \lambda_2 \: b_2,  \label{se9}\\
b_1 &=& \sqrt{\chi \; \phi} \: b_2,   \label{se9bis}
\end{eqnarray}
where the multiplicative constant $b_2$ is arbitrary and its value
depends on the initial condition (see below).
We note that $\varphi$ depends on the  parameters
 $\Pi, \delta, \chi, \phi, \Gamma_1, \Gamma_2 $ (but not $C^0$)  and
 has infinitely many roots (eigenvalues), which are real and distinct. \par

For each eigenvalue couple $ (\lambda_{1m},  \lambda_{2m}), \: m=0,1,2,...,$ satisfying (\ref{se11}), the constants $a_{1m}$, $b_{1m}$ and
$a_{2m}$ are obtained from (\ref{se9}),  (\ref{se9bis}) and (\ref{se8})
respectively, and thus the corresponding  eigenfunctions
$X_{1m}$ and $X_{2m}$ defined in (\ref{sec31}) are computed as:
\begin{eqnarray}
 X_{1m}&=&b_{2m} \tilde X_{1m}= b_{2m} \left[\tilde a_{1m}
\cos(\lambda_{1m} x) +  \tilde b_{1m} \sin(\lambda_{1m} x) \right],  \label{exp1}\\
 X_{2m}&=&b_{2m} \tilde X_{2m}=  b_{2m} \left[\tilde a_{2m} \cos(\lambda_{2m} x)
+ \sin(\lambda_{2m} x)\right],  \label{exp2}
\end{eqnarray}
where the tilde indicates a variable which has been scaled by $b_{2m}$. \\
Furthermore, the corresponding time-variable functions $G_{1m}$ and $G_{2m}$ defined
 by equations (\ref{sec3}) are computed as:
\be
G_{1m}=\exp(-\lambda_{1m}^2 t), \qquad\qquad\qquad\qquad
G_{2m}=\exp\left(- {\chi \over \phi} \lambda_{2m}^2 t \right).
\ee
($G_{1m}=G_{2m}$). Finally, the complete solution of the problem is given by a linear superposition
 of the fundamental solutions (\ref{eq1}) in the form:
\begin{align}
&c_1(x,t)=\sum_{m=1}^{\infty} A_m \tilde X_{1m}(x) \: \exp(-\lambda_{1m}^2 t),  \nonumber  \\
&c_2(x,t)=\sum_{m=1}^{\infty} A_m \tilde X_{2m}(x) \: \exp\left(- {\chi \over \phi} \lambda_{2m}^2 t \right),
\label{sl33}
\end{align}
where the arbitrary constants $A_m (= b_{2m})$  are determined through the
initial conditions (\ref{7'}).
The damping factors  $\exp(-\lambda_{1m}^2 t)$ and
 $\exp\left(- \ds{\chi \over \phi} \lambda_{2m}^2 t \right)$,  $m=1,2,...,$
 measure the attenuation of the various terms
in summations (\ref{sl33}). Because of the fast exponential convergence,
the series (\ref{sl33}) will be truncated at a
finite number of terms, in
accordance with the accuracy desired at the time of interest.
Since $\max_x | A_m \tilde X_{im}(x) | < 1$ for any $i = 1,2$, $m > 1$, to reach an accuracy of $10^{-r}$, it is sufficient to consider a finite series summation up to the index $j > 1$
such that
\bdm
\lambda_{1 j}> \sqrt{r \ln 10 \over t}
\edm
and the series is truncated at the first $j$ terms. A value of
$j = 30$ is considered for all times in the simulations.

\bigskip\bigskip
\noindent \underline{Application of the initial condition} \\
By evaluating  (\ref{sl33}) at $t=0$ and multiplying it by $\tilde X_{1n}, \tilde X_{2n}$,
after integration  we obtain:
\be
\int\limits_{-\delta}^0 \sum A_m \tilde X_{1m} \tilde X_{1n} \, dx =
 \int\limits_{-\delta}^0 \tilde X_{1n} \, dx,  \;\;\;\qquad n=1,2,..., \label{int2}
\ee
and
\be
\int\limits_{0}^{1-\delta} \sum A_m \tilde X_{2m} \tilde X_{2n} \, dx = C^0 \int\limits_{0}^{1-\delta} \tilde X_{2n} \, dx,    \qquad n=1,2,... \label{int3}
\ee
By combining equations (\ref{int2}) and (\ref{int3}) and by using
the orthogonality property of $(X_{1m}, X_{2m})$ \cite{pontrelli3} :
\be
A_m \left( \int\limits_{-\delta}^0 \tilde X_{1m}^2 \, dx +
\phi \int\limits_0^{1-\delta} \tilde X_{2m}^2 \, dx  \right)=
 \int\limits_{-\delta}^0 \tilde X_{1m}  \, dx +
\phi C^0 \int\limits_{0}^{1-\delta} \tilde X_{2m}  \, dx,  \label{int4}
\ee

we have:
\be
A_m = {  \int\limits_{-\delta}^0 \tilde X_{1m} \, dx +
\phi C^0 \int\limits_0^{1-\delta} \tilde X_{2m} \, dx   \over
 \int\limits_{-\delta}^0 \tilde X_{1m}^2 \, dx +
\phi \int\limits_0^{1-\delta} \tilde X_{2m}^2 \, dx
}.  \label{int5}
\ee

\subsection{Computing Mass}
The total mass of drug  at any time can be evaluated by integrating the drug concentrations in each layer over their respective spatial domain.
If we normalize the total mass  by its initial value,
then the non-dimensional total mass of drug in the coating
is given by
\begin{equation}
M(t)= \frac{1}{\delta+\left(1-\delta\right)\phi C^0}\left[\INT_{-\delta}^{0} c_1(x,t) dx+\phi \INT_{0}^{1-\delta}c_2(x,t) dx\right].
\end{equation}
Letting $\theta_i$ represent the non-dimensional mass of drug in each layer as a fraction of the total mass, we then have:
\begin{align}
&\theta_1(t)=\frac{1}{\delta+\left(1-\delta\right)\phi C^0 }\INT_{-\delta}^{0} c_1(x,t) dx \nonumber \\
&=\frac{1}{\delta+\left(1-\delta\right)\phi C^0 }\sum_{m=1}^{\infty} A_m \left(  {a_{1m} \sin( \lambda_{1m} \delta) +b_{1m} \cos( \lambda_{1m} \delta) -
b_{1m} \over  \lambda_{1m}} \right) \exp(-\lambda_{1m}^2 t), \label{perc1} \\
&\theta_2(t)=\frac{\phi}{\delta+\left(1-\delta\right)\phi C^0 }\INT_{0}^{1-\delta}c_2(x,t) dx  \nonumber \\
&=\frac{\phi}{\delta+\left(1-\delta\right)\phi C^0 } \sum_{m=1}^{\infty} A_m \left(  {a_{2m} \sin( \lambda_{2m} (1-\delta)) - \cos( \lambda_{2m} (1-\delta)) +1
 \over  \lambda_{2m}} \right) \exp\left(-{\chi \over \phi}
\lambda_{2m}^2 t \right)  \label{perc2}
\end{align}
where $c_i$ ($i=1,2$) are given by (\ref{sl33}).
It is then straightforward to evaluate the total non-dimensional mass of drug in
the coating as $M=\theta_1+\theta_2$ and the cumulative fraction of drug release,
$M_{frac}$ (the \textit{release profile}) as $M_{frac}=1-\left(\theta_1+\theta_2 \right)$.

The depletion of the drug in coating as a result of the release process is governed by an
exponential decay as in the above equations. The analytical solution indicates that a complete release is reached
only asymptotically and equations (\ref{perc1})--(\ref{perc2}) allow one to estimate the release time $T_r$, within a given
tolerance $\epsilon$, through:
\bdm
 \theta_1(T_r)+\theta_2(T_r) \leq \epsilon.
\edm
The smallest eigenvalue $\lambda_1^{min}= \ds\sqrt{\chi \over \phi}  \lambda_2^{min}$ relates to the
dominant damping factor in the series (\ref{sl33}), (\ref{perc1})-(\ref{perc2}). Comparing $\lambda_1^{min}$ between different parameter regimes provides an indication of the relative rate of release (see Table 2).
For the particular parameter regime of interest and a given initial mass per cross-sectional area $M^0$, the initial loading concentrations are calculated through:
\begin{eqnarray}
c_1^0&=&\frac{M^0}{ L \phi_1\left(\delta + \phi C^0\left(1-\delta\right)\right)} \nonumber\\
c_2^0&=&c_1^0 C^0.\label{back-calculate}
\end{eqnarray}


\subsection{Special Case}
We note that in the special case where the microstructural properties of the layers are identical ($\chi=\phi=1$) then we can obtain an analytical solution.  In this case, the eigenvalues $\lambda_1=\lambda_2=\lambda$, say, and are  obtained by solving \[\cos\left(\lambda\right)=0.\]  The difference between the solutions $c_1$ and $c_2$ then arises only through $A_m$, which is calculated using the initial condition.  It can be shown that the solution in this case is
\begin{eqnarray}
\label{sc1}
c(x,t)=\frac{4}{\pi}\sum_{n=1}^{\infty}\frac{\left(-1\right)^n \cos\left(\frac{\pi(2n-1)(x+\delta)}{2}\right)\left\{(-1)^{n+1}\sin\left(\frac{\pi(2n-1)\delta}{2}\right)\left(C^0-1\right)-C^0\right\} \exp{\left(-\frac{\pi^2\left(2n-1\right)^2 t}{4}\right)}}{2n-1},&& \nonumber \\
-\delta<x<1-\delta,&& \nonumber \\
\end{eqnarray}
and the release profile is readily computed as:
\begin{eqnarray}
\label{sc2}
M_{frac}=1+\frac{8}{\pi^2\left(\delta+(1-\delta)C^0\right)}\sum_{n=1}^{\infty}\frac{\left\{(-1)^{n+1}\sin\left(\frac{\pi(2n-1)\delta}{2}\right)\left(C^0-1\right)-C^0\right\}\exp{\left(-\frac{\pi^2\left(2n-1\right)^2 t}{4}\right)}}{(2n-1)^2}. \nonumber \\
\end{eqnarray}

\section{Results and Discussion}
\setcounter{equation}{0}
In all simulations, we consider the typical boundary conditions $\Gamma_1=0$ and $\Gamma_2\rightarrow\infty$.  This situation is representative of the
most common case where the coating is in contact with an impermeable material  (e.g. metal structure of the device) on one side and is exposed to an infinite sink at the other side, where the drug is washed away instantaneously.  It is usual for infinite sink conditions to be maintained during in-vitro drug release experiments.  At the interface between the two layers, we choose $\Pi\rightarrow \infty$ to reflect `unhindered' transport.

The free diffusion coefficient of molecules in liquids $D^w$ is typically of the order of $10^{-9}\; m^2\; s^{-1}$ \cite{cussler}.  By definition, $0<\phi_i^e<1$.  However, extremely low ($\phi_i^e<0.1$) and extremely high ($\phi_i^e >0.9$) porosities would likely result in  drug loading and mechanical constraints, respectively.  A typical range of tortuosity values is $1<\tau_i<6$, although values as high as 10 have been reported \cite{cussler}. Taken together, we expect that the effect of the microstructure in each layer is to result in an effective diffusion coefficient at most two orders of magnitude smaller than the free diffusion coefficient in water.  We note that drug diffusion coefficients in some polymers have been reported to be as low as $10^{-17}\; m^2\;s^{-1}.$  However, it should be noted that these are usually {\em apparent} diffusion coefficients which likely incorporate other effects such as absorption and desorption (possibly in addition to the microstructure effects that we consider here).  The effect of such processes can be to reduce the overall diffusion coefficient by several orders of magnitude. In all simulations we fix $D_1^e=5\cdot 10^{-11}$  and  $\phi_1=0.6$ and consider the effects of varying the microstructure of each layer by varying $\phi$ and $\chi$.

Since the purpose of this study is to establish the benefits of replacing a single layer with a two-layer coating of identical total thickness $L$ and initial drug mass, we fix $L=10^{-4}m$ in all simulations.
In reality, of course, the values of $D_1^e$, $\phi_1$ and $L$ will vary depending on the particular application.  Since our focus is to investigate the effect of the results on varying the  \textit{ratio} between the parameters of each layer, we have decided to choose broadly typical values, whilst acknowledging that this will not cover all cases.
In all of our simulations the initial non-dimensional mass is 1. We choose not to mathematically implement a fixed dimensional mass. As a consequence, the initial dimensional loading concentrations $c_1^0$ and $c_2^0$ are to be back-calculated using (\ref{back-calculate}) such that the desired initial dimensional mass is achieved.

\bigskip


\subsection{Baseline model}
To assess the effect on drug release of variations in system parameters, we preliminarily assume that layer 1 and layer 2 have identical microstructural parameters ($\chi=\phi=1$) and equal initial drug concentrations ($C^0=1$): in this case we can use the analytical solutions given by (\ref{sc1}-\ref{sc2}).
 The result is that our baseline model (see Table 1) essentially reduces to a single layer system (the solution is independent of the choice of $\delta$).  The resulting non-dimensional parameters are $\chi=\phi=C^0=1$.

\begin{table}[h]
 \footnotesize
 \begin{center}
\begin{tabular}{|c|c|}
\hline
 Parameter  & Value (layer 1 \quad $-$ \quad layer 2)    \\
\hline
\hline
$D_i^e (\rm{m^2  s^{-1}})$   & $5 \cdot 10^{-11} \quad - \quad 5 \cdot 10^{-11}  $         \\
 \hline
$L_i (\rm{m})$   & $5 \cdot 10^{-5} \quad - \quad  5 \cdot 10^{-5} $     \\
 \hline
$\phi_{i}$  &  $  0.6 \quad - \quad 0.6   $ \\
 \hline
  $K_i$  &   $0 \quad -  \quad  10^{10} (^{*}) $ \\
 \hline
   $P (\rm{m s^{-1}})$  &   $10^{10} (^{*}) $  \\
 \hline
\end{tabular}
\caption{Reference dimensional parameter values used in the baseline simulations.  The two layers have the same physical parameters and unhindered transport between the layers, and so they are equivalent to one layer. $(^{*})$ In reality, we wish to impose $K_2, P \rightarrow \infty$, however,  for the purposes of the numerical simulations it was found that the value $10^{10}$ was sufficient to represent this case. }
\end{center}
\end{table}

\begin{figure}
\centering\scalebox{0.4}{\includegraphics{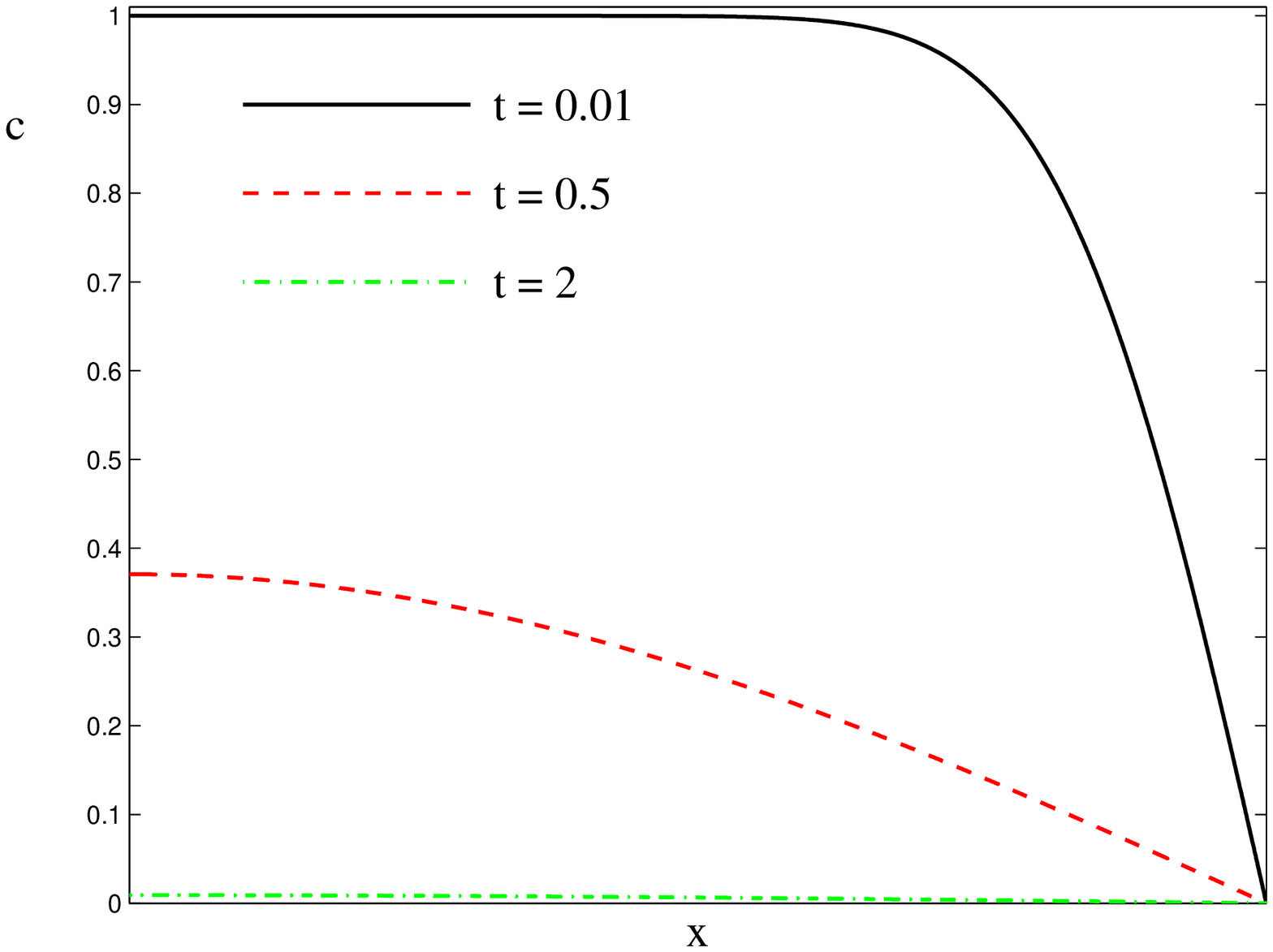}}
\hspace{3mm}
\centering\scalebox{0.4}{\includegraphics{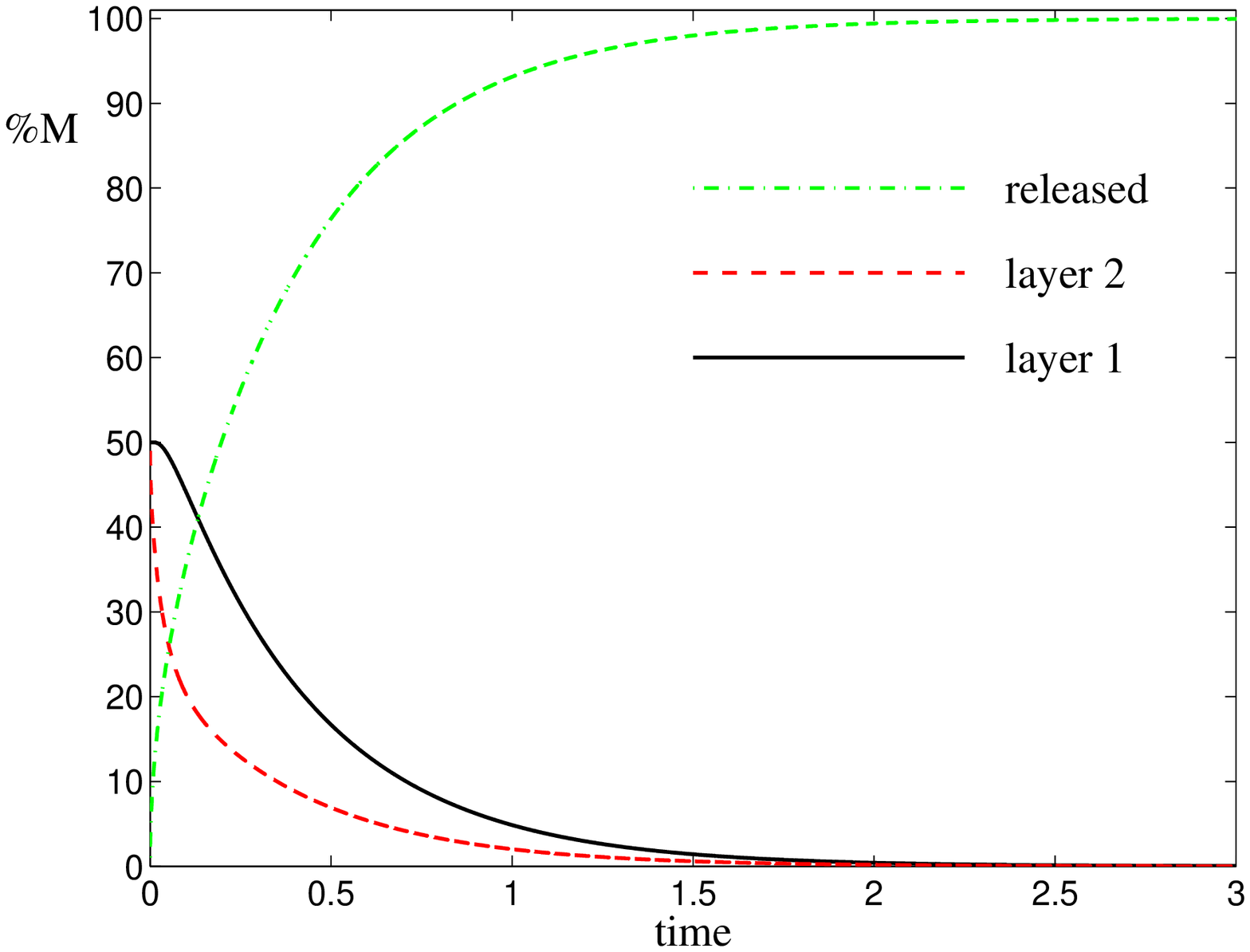}}
\caption{Baseline case: the non-dimensional concentration profiles at three times and percentage of drug mass released versus time (one layer, Table 1). Because of the perfect contact at the interface ($\Pi \rightarrow \infty$),  the concentration curves
results are insensitive to the location of the interface  (left). The mass percentages in the individual layers do vary with $\delta$, but the release curves (green) do not (right).}
\label{fig4}
\end{figure}

  In Figure \ref{fig4} we display the results of the baseline case, where each layer has identical initial drug loading and microstructure, so that we effectively have a single layer. As a result of the infinite sink boundary condition at the release medium, drug is rapidly released from layer 2 in the early stages, whilst there is a small delay before drug concentrations in layer 1 drop from their initial value.  Drug release from layer 1 proceeds at a slower rate than in layer 2, and therefore there is a difference in both the shape and the duration of release in each layer.  All of the drug has been released from the system by approximately $t=3$ (non-dimensional time).

\subsection{Sensitivity analysis}
We are ultimately interested in quantifying the effect on drug release of having two separate layers (as opposed to a single layer) with different microstructure and drug loading parameters. Therefore, it is of interest to vary the parameters, one at a time,  around the baseline values and to compare the resulting drug release profiles.  We consider three cases (see Table 2): in Study 1 we assess the effect of varying $\chi$, whilst in Study 2 and Study 3 we vary $C^0$ and $\phi$, respectively.   In each case we consider three values of $\delta$.

\begin{table}[h]
 \footnotesize
 \begin{center}
\begin{tabular}{|c|c|c|c|c|}
\hline
Study   & $\chi$& $C^0$& $\phi$ & $\lambda_1^{min} (\delta=0.5)$  \\
\hline
\hline
Baseline & 1 & 1& 1 & 1.57 \\
\hline\hline
&0.5&1&1&  1.16 \\
\cline{2-5}
\raisebox{1.5mm}{1}&2&1&1& 2.03 \\
\hline \hline
&1&0&1 &1.57\\
\cline{2-5}
\raisebox{1.5mm}{2}&1&5&1& 1.57\\
\hline \hline
&1&1&2/3 & 1.61 \\
\cline{2-5}
\raisebox{1.5mm}{3}&1&1&3/2 & 1.50 \\
\hline
\end{tabular}
\end{center}
\caption{Range of non-dimensional parameters simulated. In each Study, three
values of $\delta (0.2, 0.5, 0.8)$  were used. $\lambda_1^{min}$ is an indicator of the release time.}
\end{table}

\bigskip
\underline{Study 1: effect of  varying microstructure ratio $\chi$ }

\begin{figure}[h!]
\centering\scalebox{0.4}{\includegraphics{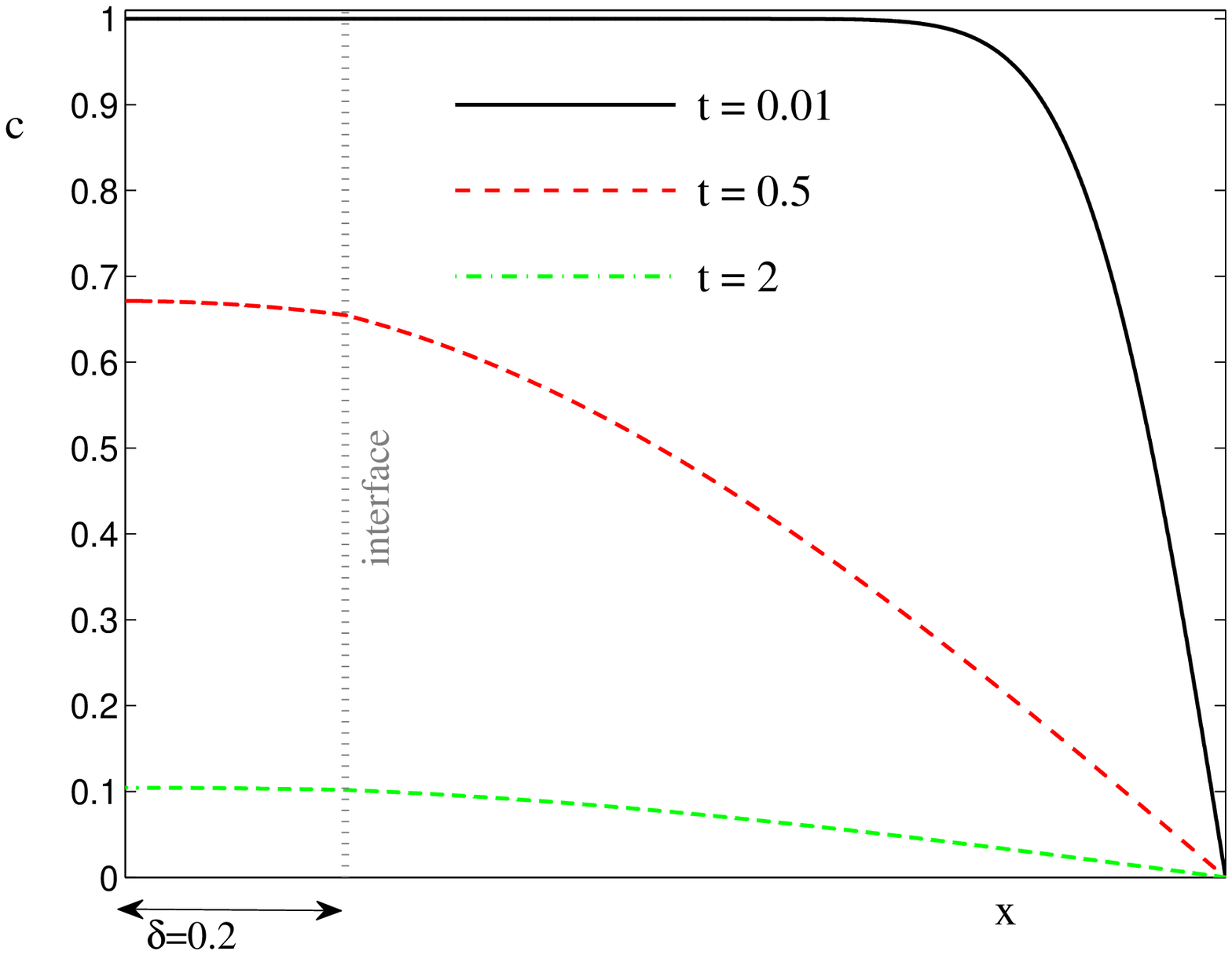}} \hspace{10mm}
\centering\scalebox{0.4}{\includegraphics{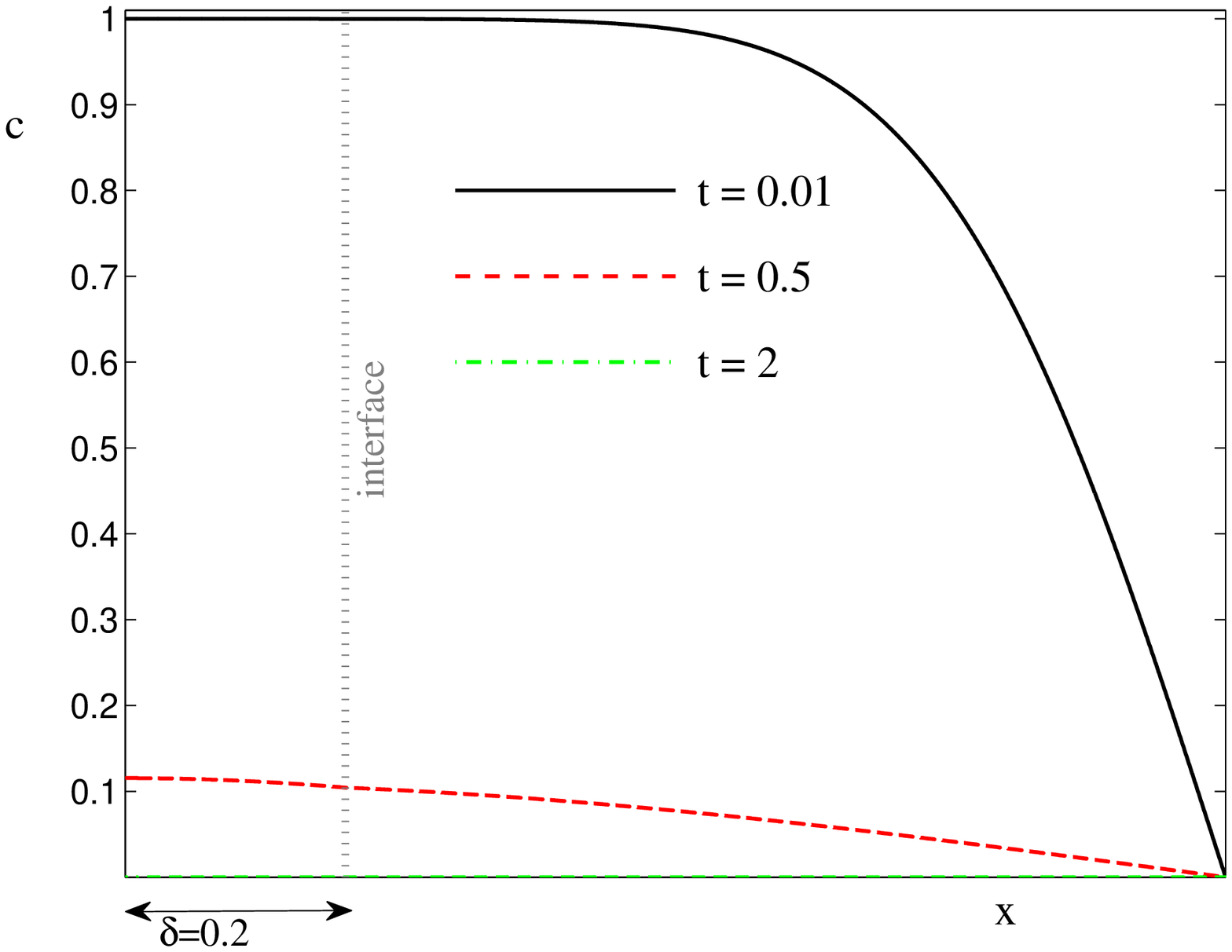}}
\vspace{5mm}\\
\centering\scalebox{0.4}{\includegraphics{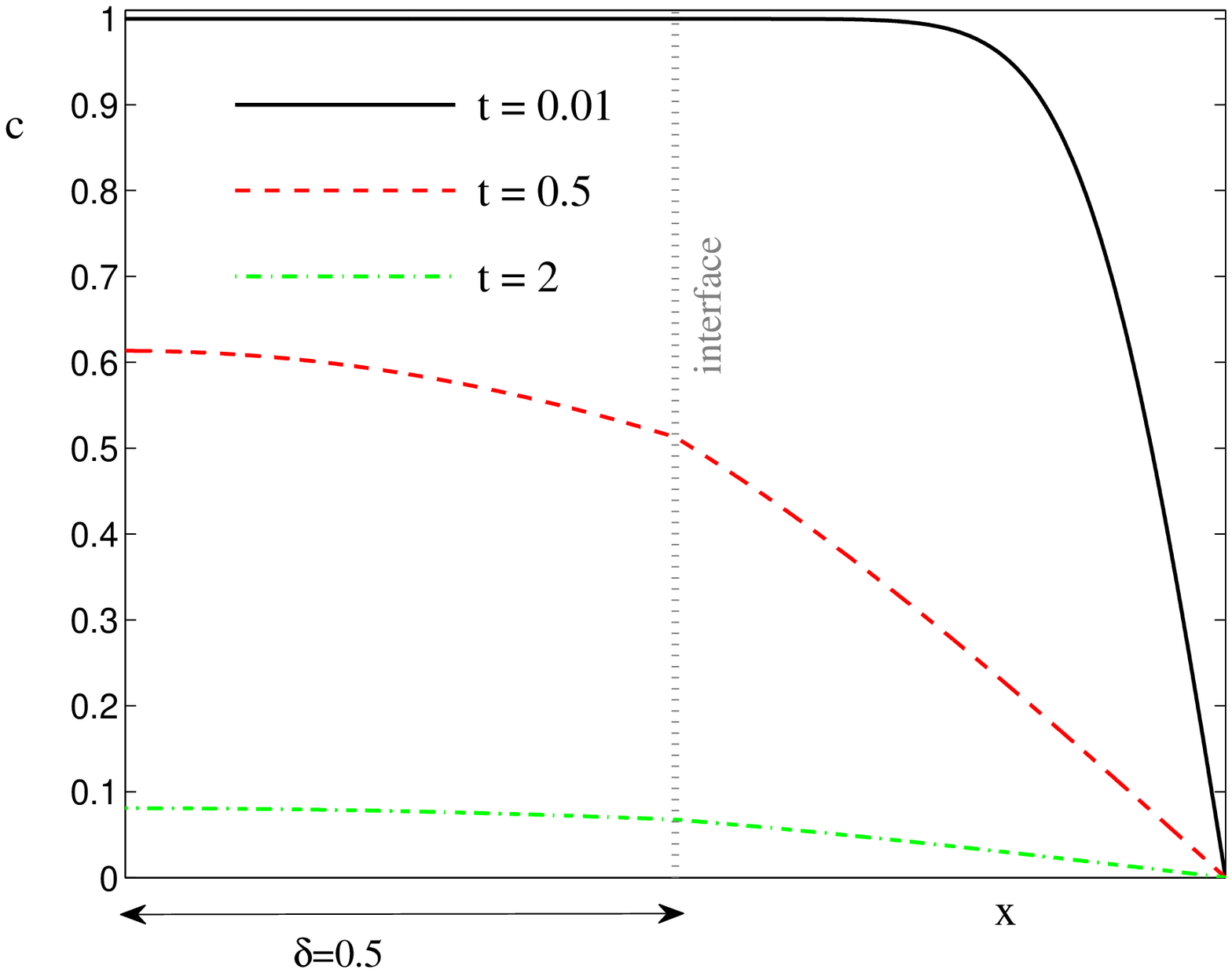}} \hspace{10mm}
\centering\scalebox{0.4}{\includegraphics{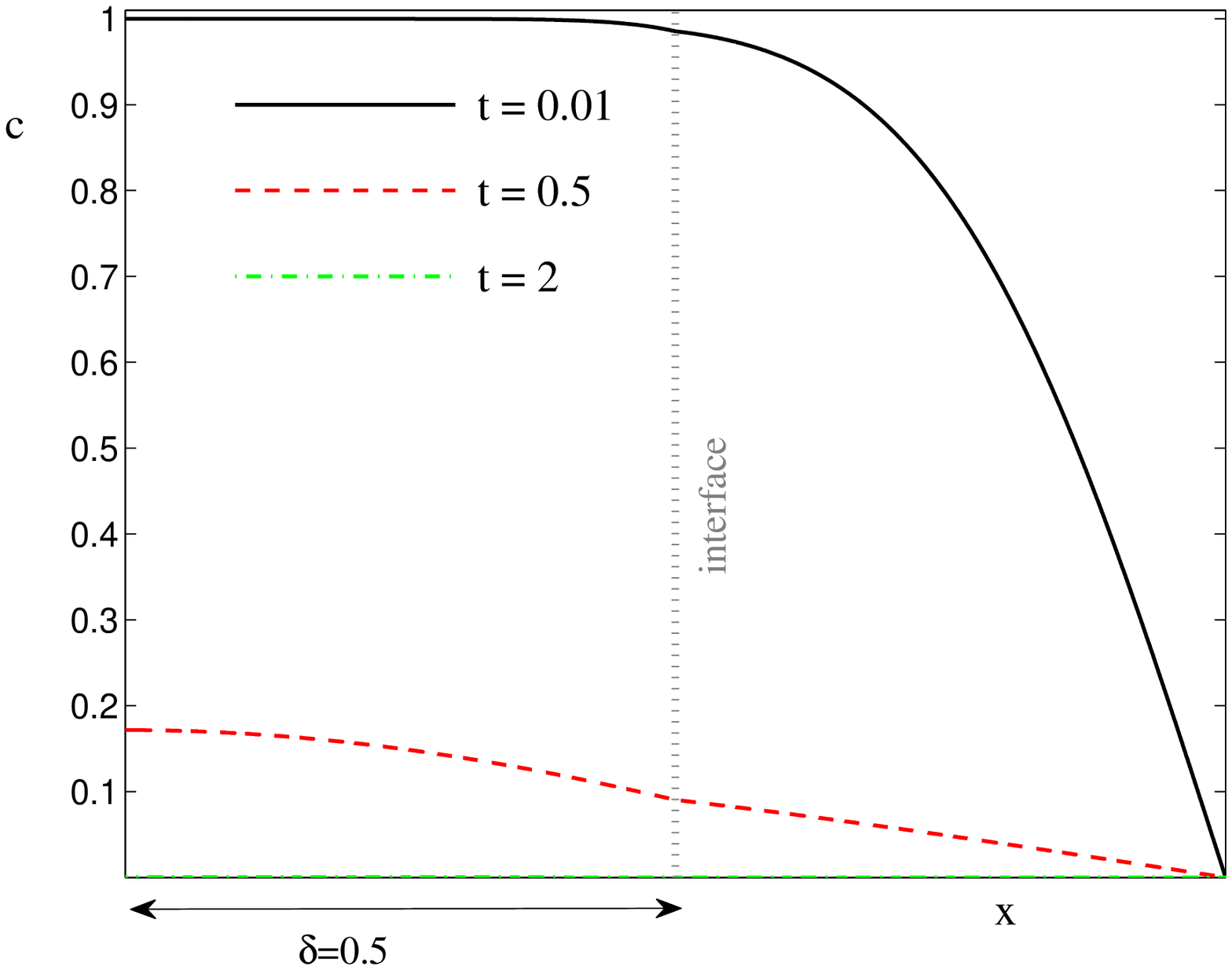}}
\vspace{5mm}\\
\centering\scalebox{0.4}{\includegraphics{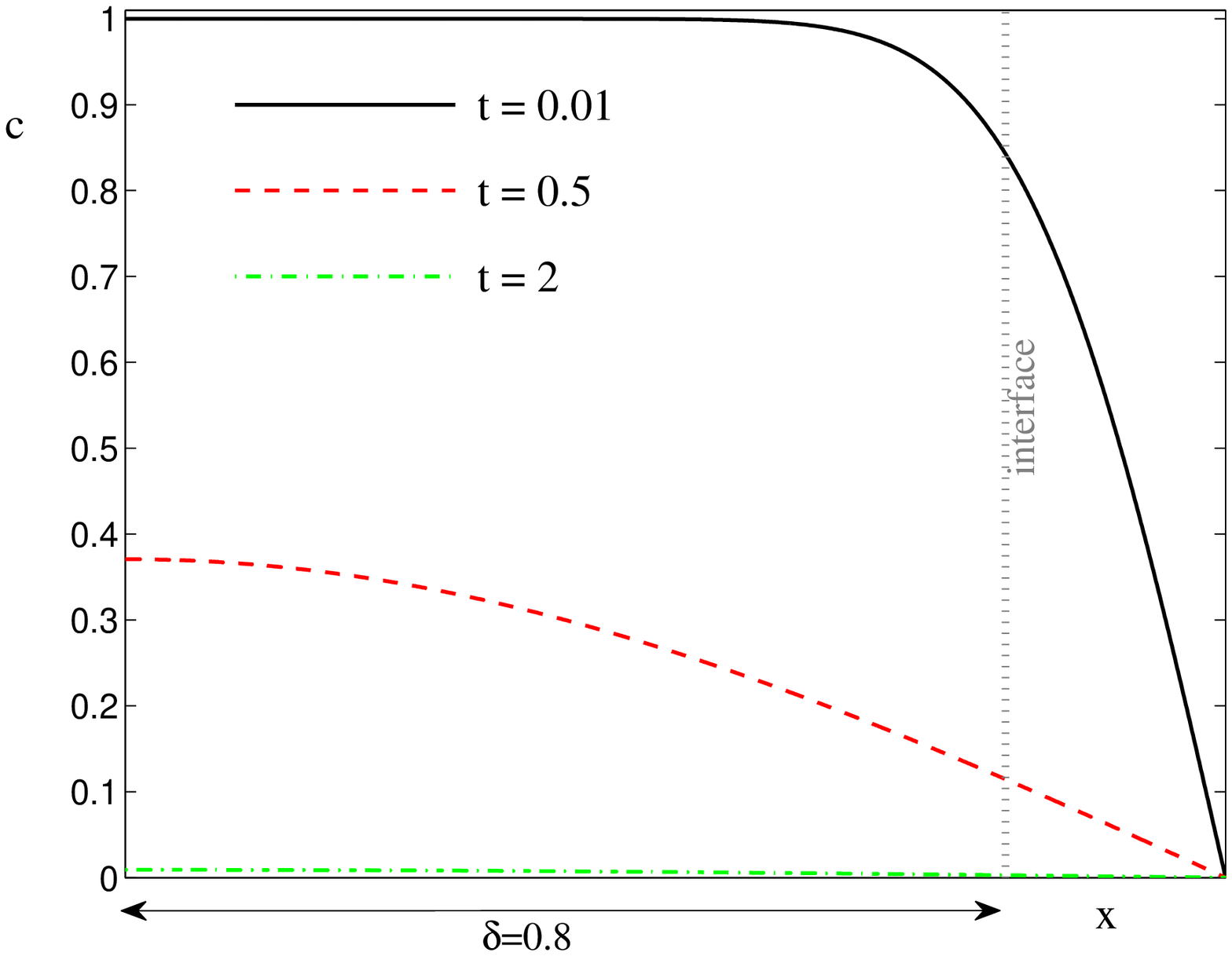}} \hspace{10mm}
\centering\scalebox{0.4}{\includegraphics{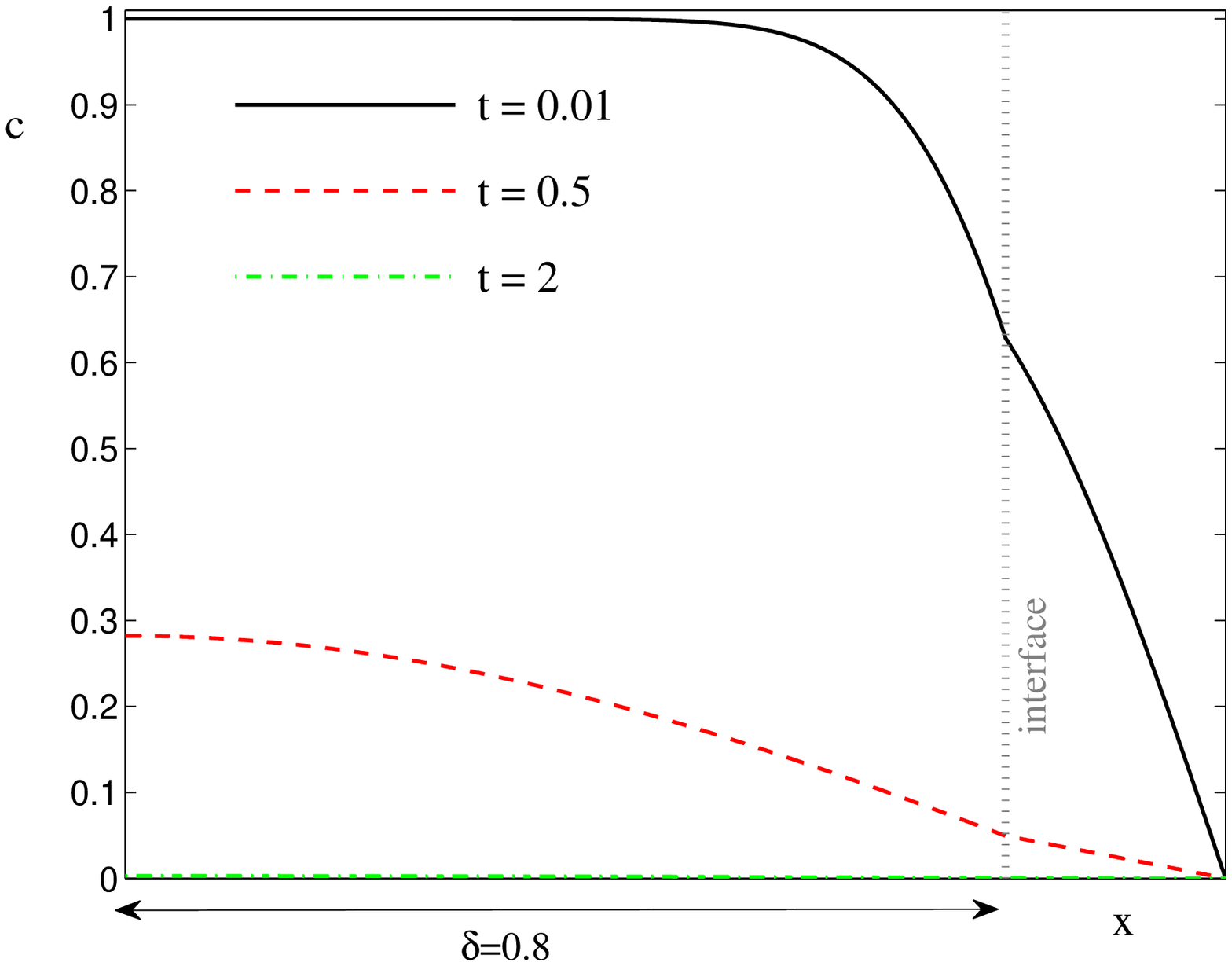}}
\caption{Non-dimensional concentration profiles for three layer thickness ratios $\delta$, with $\chi=0.5$ (left)  and $\chi=2$ (right), and the other parameters as in Table 2 (Study 1).}
\label{fig5}
\end{figure}

\begin{figure}[h!]
\centering\scalebox{0.6}{\includegraphics{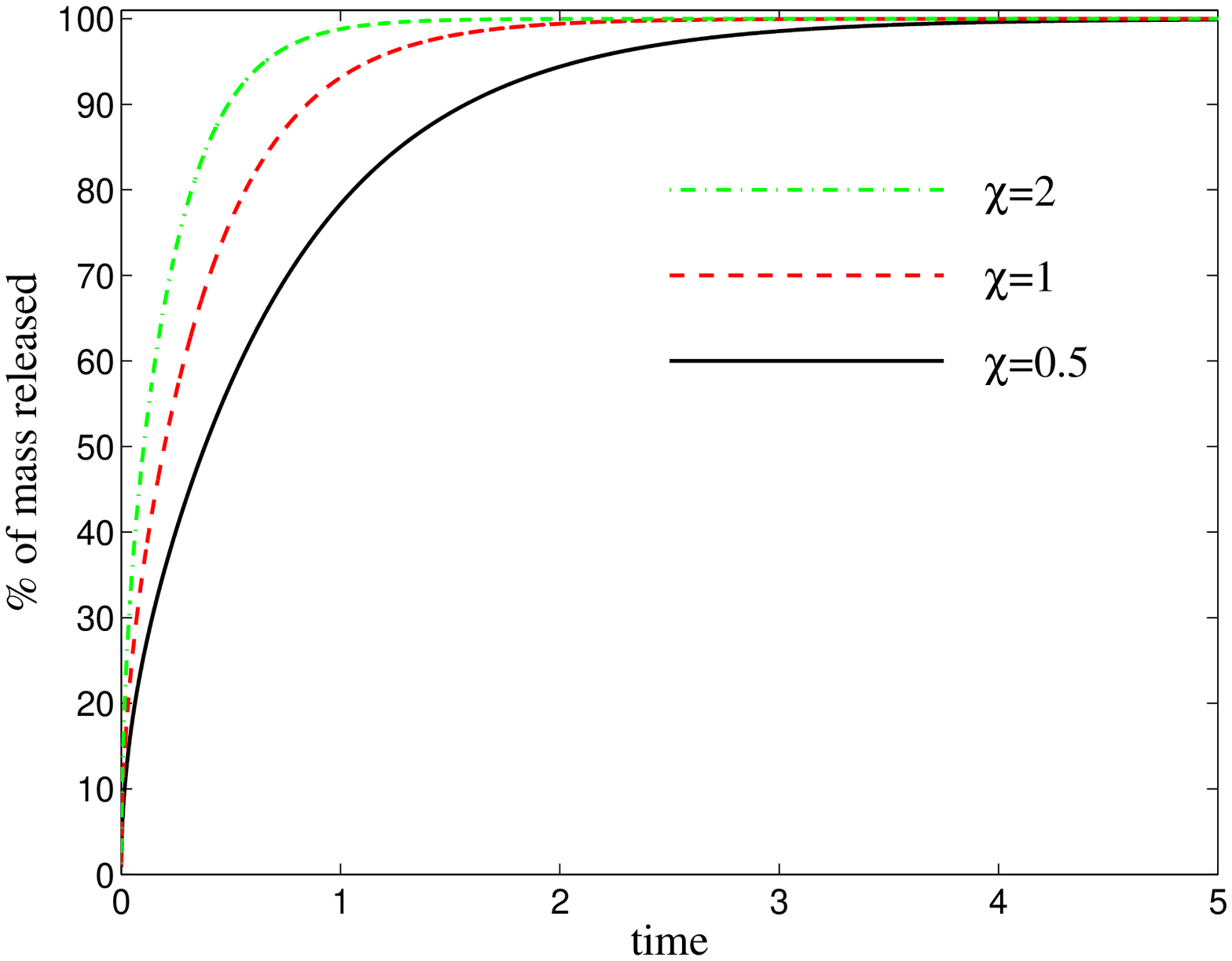}}
\caption{Comparison of release profiles at three values of $\chi$ (all the other values as in table 2 and $\delta=0.5$).  }
\label{fig6}
\end{figure}

We now assess the effect of varying the relative microstructural parameters between the two layers.  In Figure \ref{fig5}, left column, we choose $\chi$ such that the effective diffusion coefficient in layer 2 is half that of layer 1, whilst in the right column the effective diffusion coefficient is 2 times greater.  In the first case we observe that drug release from layer 1 is hindered by the lower effective diffusion coefficient in layer 2 and as a result there is a delay in drug being released from layer 1.  Despite the lower effective diffusion coefficient in layer 2, there is still a burst release as a result of the infinite sink conditions, but this effect is smaller than the baseline case.  In the second case, the faster effective diffusivity in layer 2 results not only in significantly faster drug release from layer 2, but also from layer 1 (Figure \ref{fig5}, right).  In Figure \ref{fig6} we plot the overall release profiles in  these two cases and compare with the baseline (dashed red line). We display only the case of $\delta=0.5$.  From Figure \ref{fig6}  we conclude that  the parameter $\chi$ has a strong influence on both the shape (rate of release) and the duration of release.  This is perhaps unsurprising since $\chi$ appears prominently in the exponential
damping factor (see (\ref{sl33}) and (\ref{perc2})). For $\chi=0.5$ and $\chi=2$ we observe faster release and slower release, respectively, as we increase $\delta$ (not shown).

The implication here is that, simply by varying the microstructure of the two layers, not only it is possible to alter the shape of the release profile, but it is also possible to ensure that drug is delivered over some defined period of time.  We note that although $\chi$ contains parameters relating to both the porosity and tortuosity of each layer, it is the combination of these values (rather than their individual size) which defines the release profile. For example, a value of $\chi=2$  could be obtained by doubling the effective porosity of layer 2 (in comparison with layer 1) or, by doubling the tortuosity of layer 1 (in comparison with layer 2).  Therefore, this parameter is highly important as it offers much flexibility from the manufacturing point of view.

\bigskip
\underline{Study 2: effect of varying ratio of initial concentrations $C^0$ }

\begin{figure}[h!]
\centering\scalebox{0.4}{\includegraphics{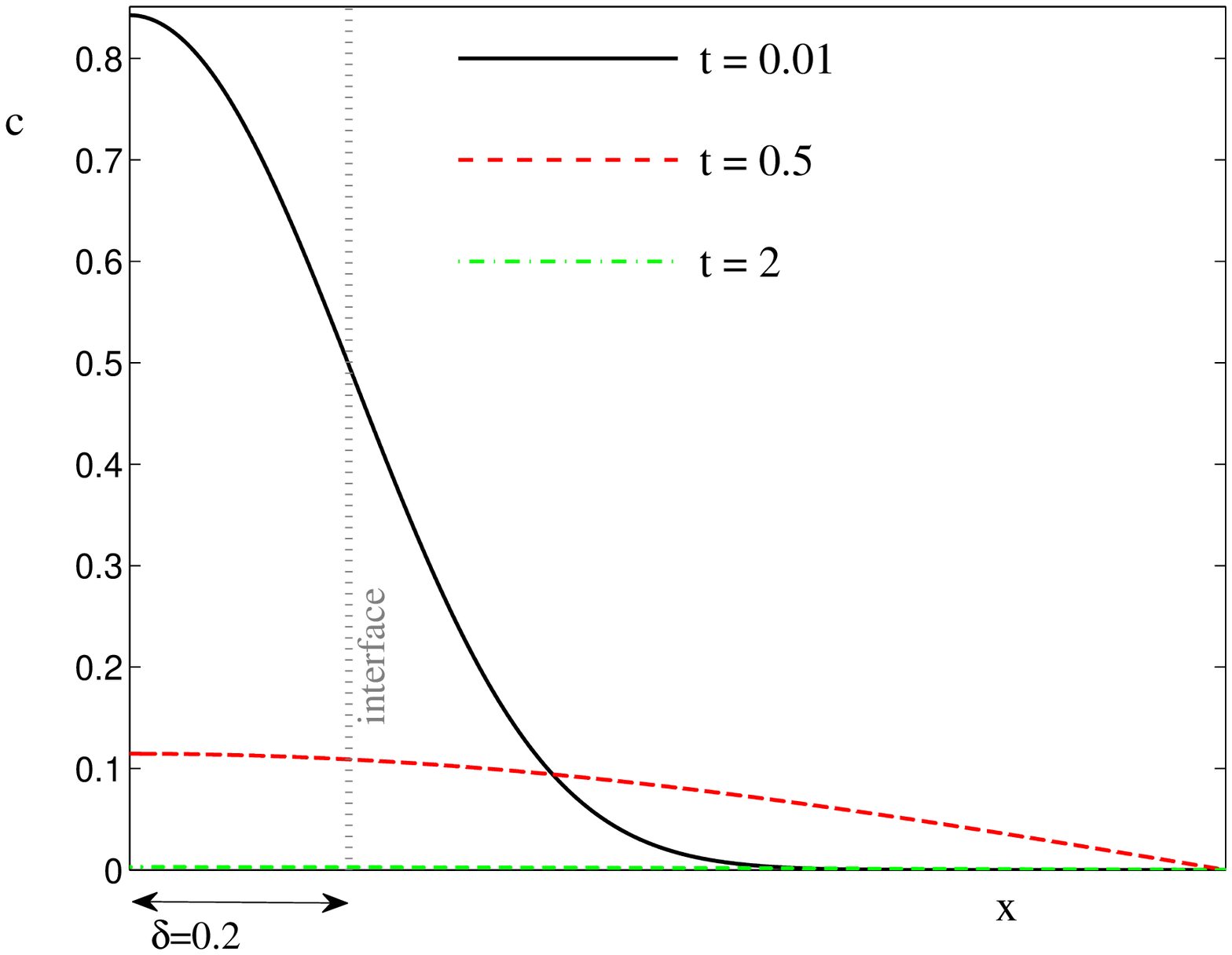}} \hspace{10mm}
\centering\scalebox{0.4}{\includegraphics{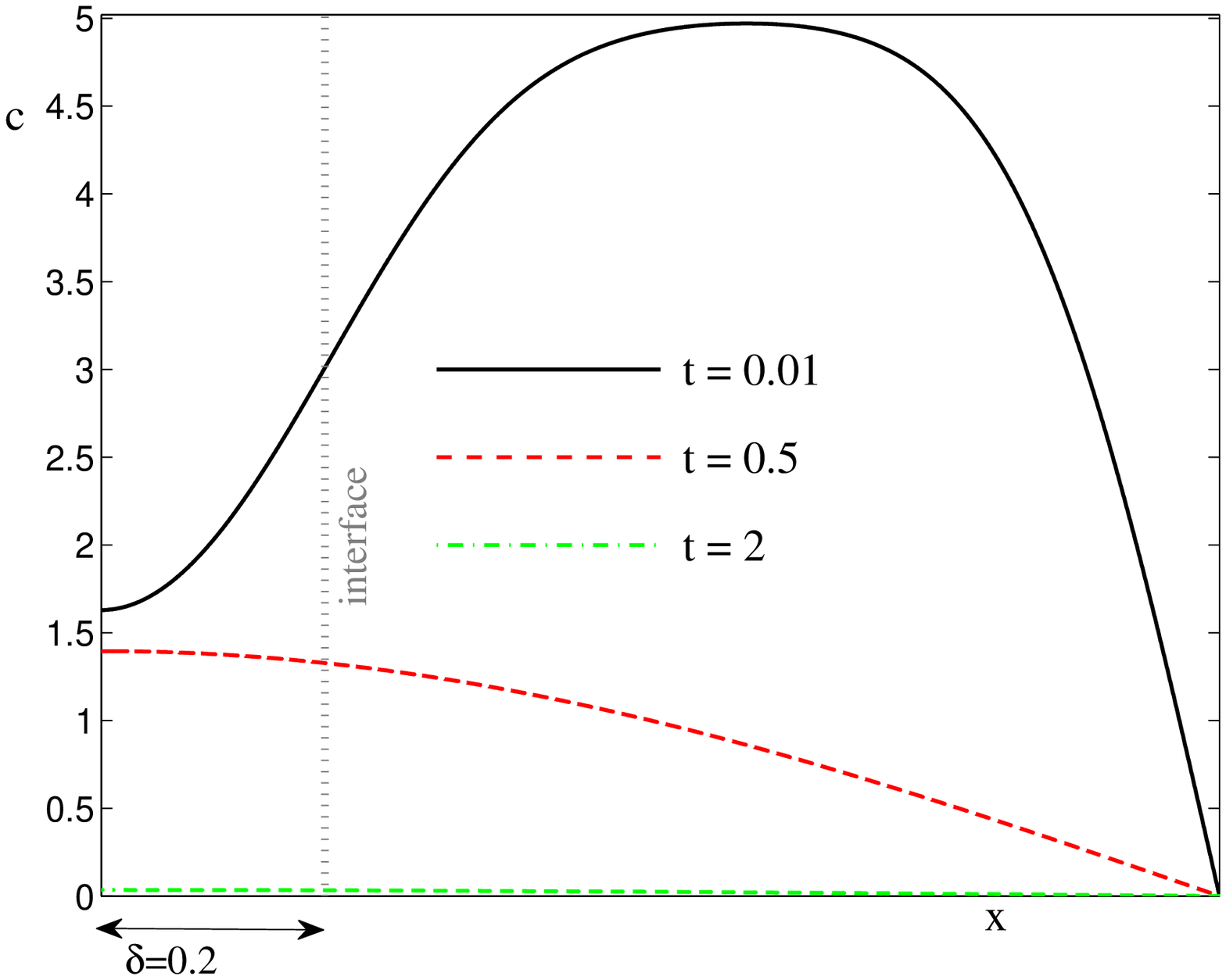}}
\vspace{5mm}\\
\centering\scalebox{0.4}{\includegraphics{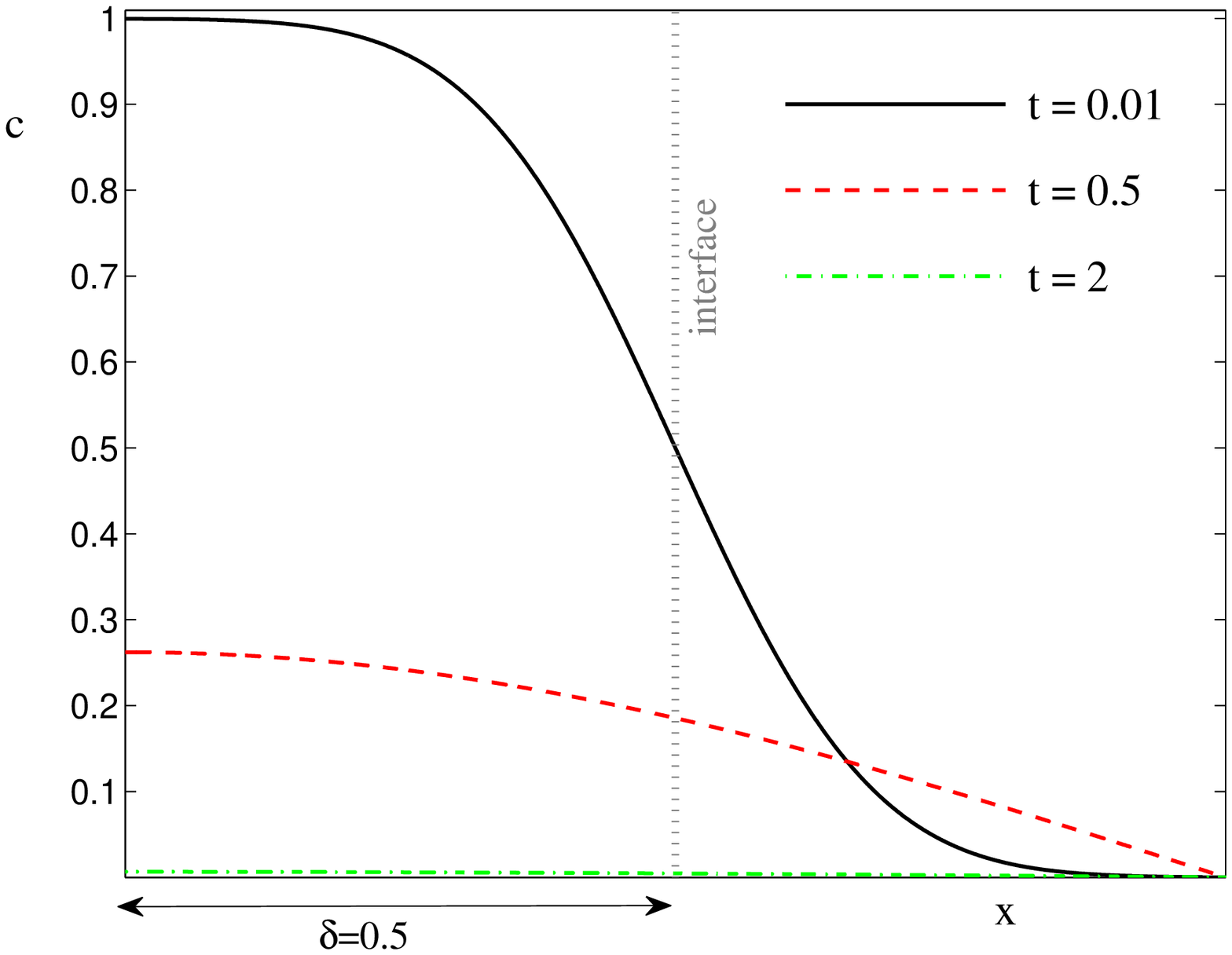}} \hspace{10mm}
\centering\scalebox{0.4}{\includegraphics{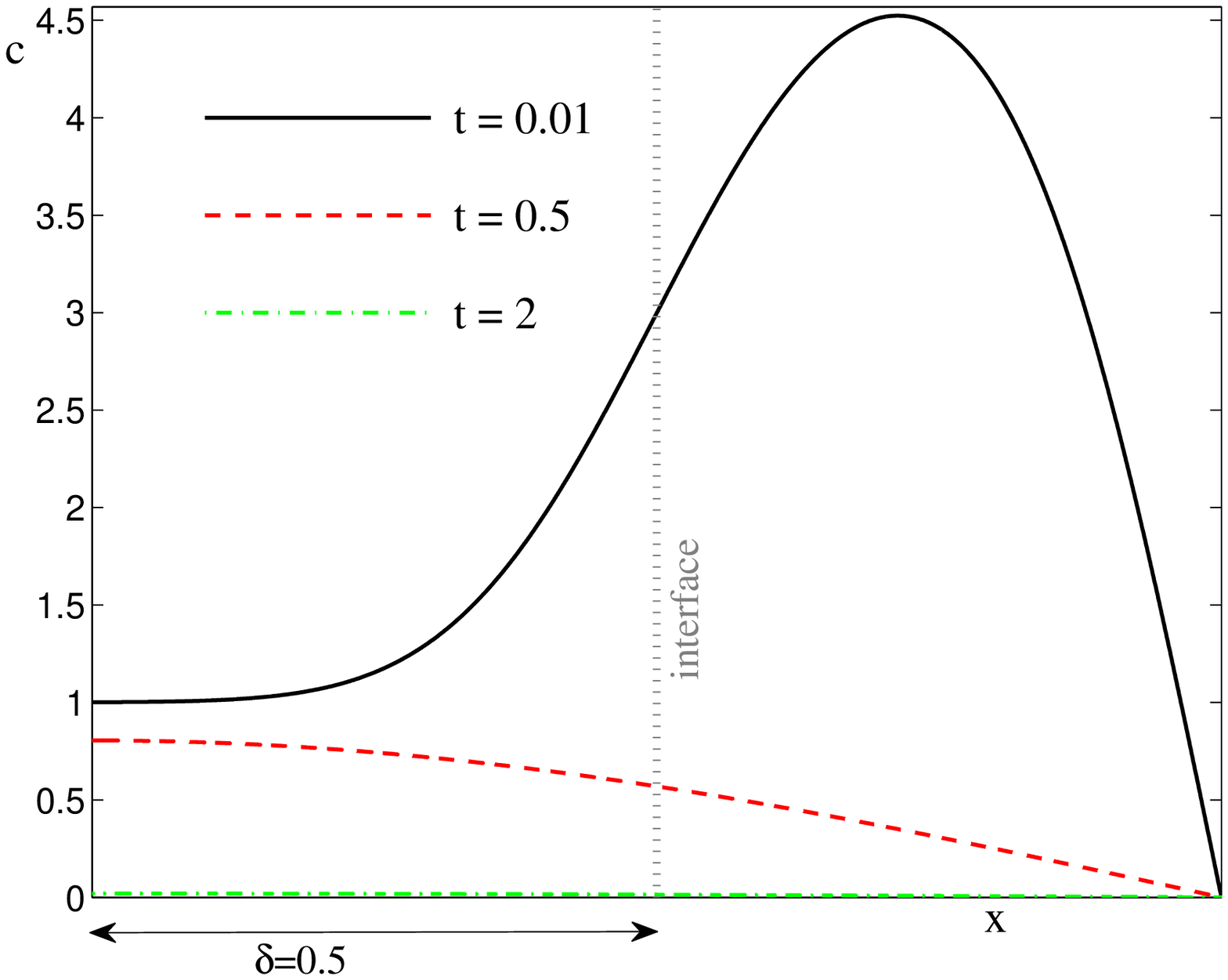}}
\vspace{5mm}\\
\centering\scalebox{0.4}{\includegraphics{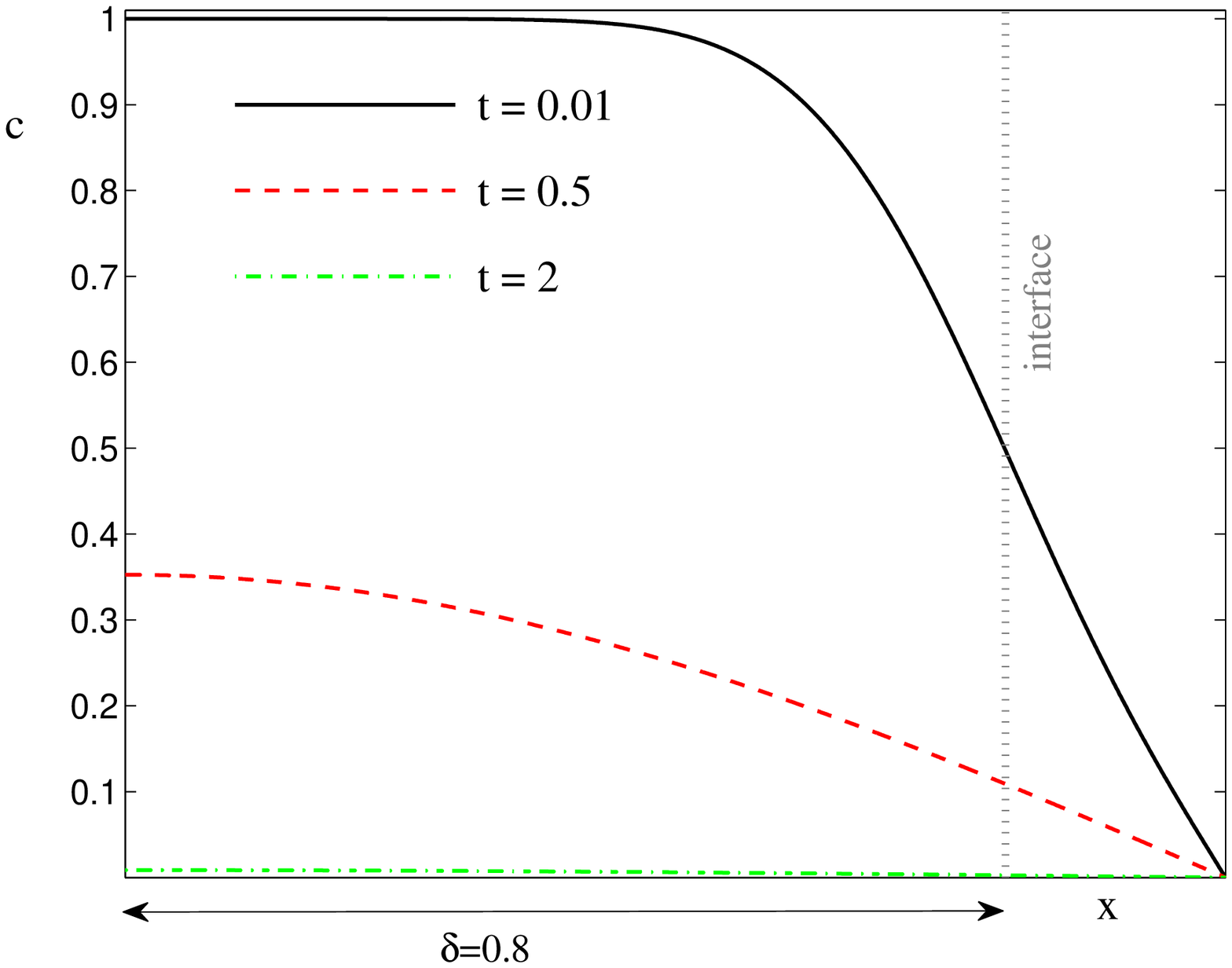}} \hspace{10mm}
\centering\scalebox{0.4}{\includegraphics{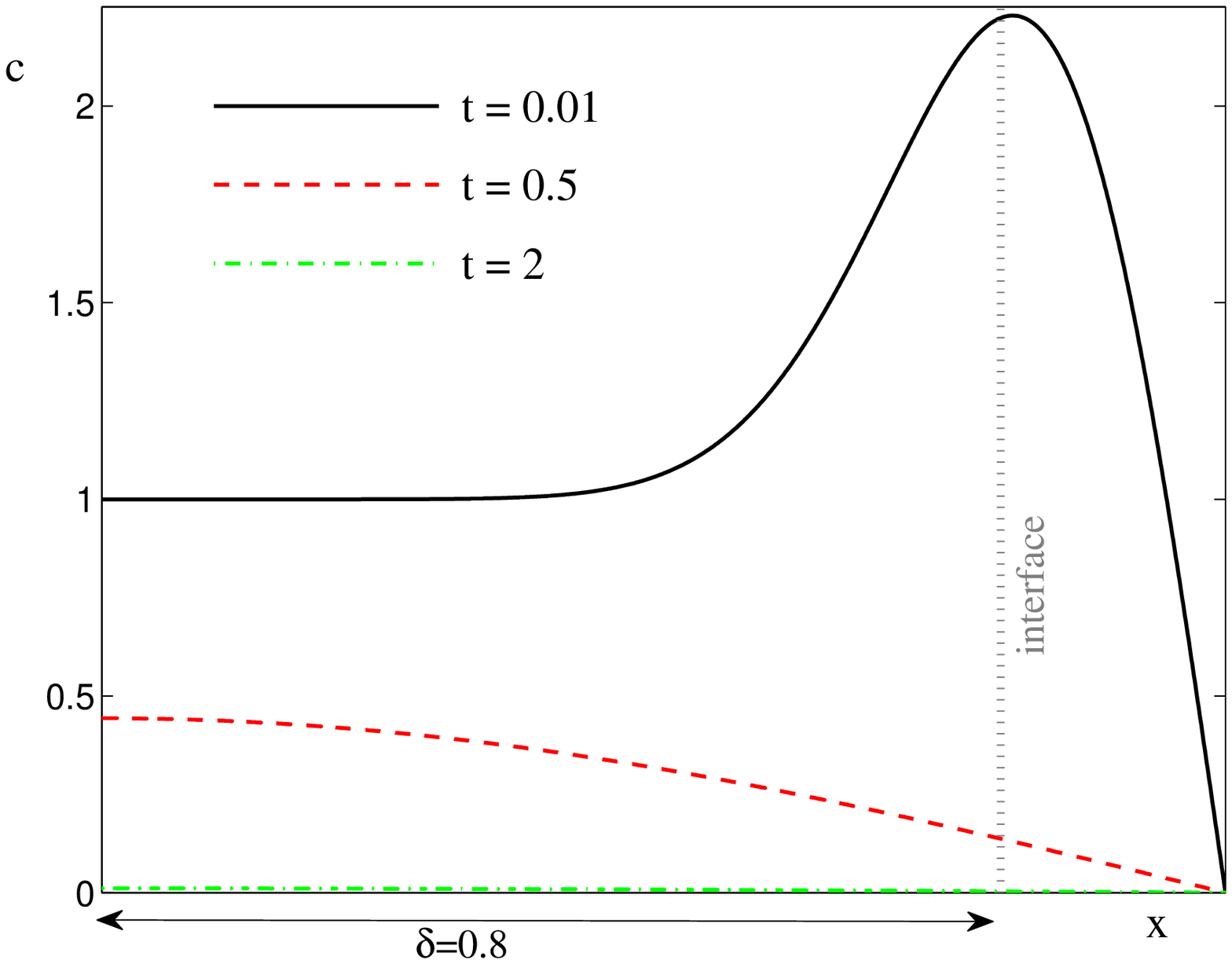}}
\caption{Non-dimensional concentration profiles for three layer thickness ratios $\delta$, with $C^0=0$ (left) and  $C^0=5$ (right) and the other parameters as in Table 2 (Study 2).  The dimensional values may be back-calculated from (\ref{back-calculate}). }
\label{fig7}
\end{figure}

We now elucidate the effect of varying the initial drug concentration between the two layers.  In the first case (Figure \ref{fig7}, left) we choose the initial drug concentration in layer 2 to be zero, whilst in the second case (Figure \ref{fig7}, right) we choose the concentration in layer 2 to be five times that of layer 1.  In the first case we observe that layer 2  is initially rapidly infiltrated with drug, before drug is subsequently released after it has traversed the thickness of the second layer.  In the second case we observe that whilst layer 2 is depleted rapidly as a result of the infinite sink condition, at early times an increase in drug concentration (and consequently drug mass) is observed in layer 1 due to the concentration gradient between the two layers (we are assuming that no drug can diffuse between the layers prior to the coating being placed in the release medium).  As layer 2 continues to be depleted of drug, eventually the concentration gradient at the interface changes direction and drug then diffuses from layer 1 into layer 2 before being released.  In each case, the value of $\delta$ has a significant impact on the concentration profile in each layer.

\begin{figure}[h!]
\centering\scalebox{0.6}{\includegraphics{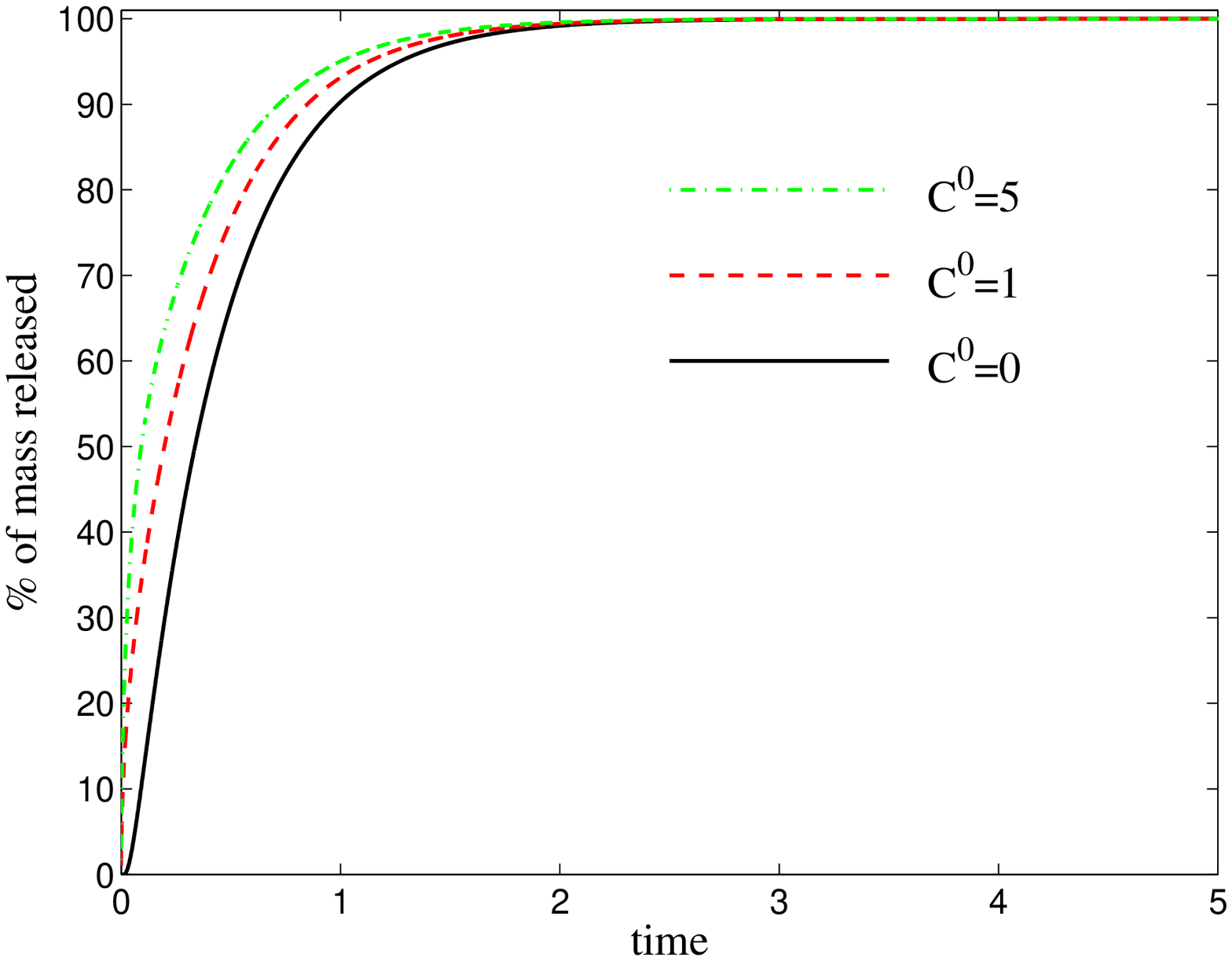}}
\caption{Comparison of release profiles at three values of $C^0$  (all the other values as in Table 2 and $\delta=0.5$).}
\label{fig8}
\end{figure}

In Figure \ref{fig8} we plot the overall release profiles in these two cases with $\delta=0.5$ and compare with the baseline (dashed red line).  For both $C^0=0$ and $C^0=5$ we observe faster release as we increase $\delta$ (not shown).  We conclude that having a drug-free second layer can delay the start of the drug-release process, which may be desirable in certain applications.  In contrast, choosing a higher initial drug concentration in the second layer can result in a larger {\em burst} of drug which also may be advantageous in other circumstances.  However, in all cases the overall duration of release results the same.   Therefore, the non-dimensional parameter $C^0$ can be used as a tuning parameter to vary the proportion of drug delivered in the initial  stages.  The inflection point at $t=0$  (Figure \ref{fig8}, $C^0=0$ - black curve) indicates a retardation time due to the filling of the second layer which is initially empty.

\bigskip
\underline{Study 3: effect of varying porosity ratio $\phi$ }
\\
\noindent By varying $\phi$ from $2/3$ to $3/2$, no significant differences are found between the concentration profiles (not shown).  The release profiles are vitually indistinguishable for high values of $\delta$, although minor differences in the profiles are observed as $\delta$ is reduced (not shown).  These results reinforce the idea that it is the effective porosity in each layer $\phi_i^e$ that drives the drug release, rather than the overall porosity $\phi_i$.
\bigskip

Having studied separately the influence of the individual parameters,
we note that  a combination of the above cases  should be considered
in order to meet the precise manufacturing requirements or with the
aim of optimising some quantity.  For example, if the objective is to slow down the release, then it appears that the simultaneous
occurrence of the two cases $\chi <1,\; C^0 =0$ will boost this property: in particular,
a configuration with a
lower effective porosity in layer 2 faced with one of higher effective porosity in layer 1 acts more
favourably to achieve this goal: the time scale for release from layer 2 is increased, and layer 1 acts as a reservoir
that continuously supplies drug during elution.

\section{Conclusions}

In this paper we have presented a mathematical model of drug diffusion through two adjacent porous layers
and we have carried out a systematic study of the effect on drug release of changes to system parameters.  Our results indicate that the contrast in properties of the two layers can be used as a means of better controlling the release, and that the quantity of drug delivered in the early stages can be modulated by varying
the distribution of drug across the layers. We conclude that microstructural and loading
differences between variable porosity coating layers can be utilized to tune the properties of the coating
materials to obtain the desired drug release profile for a given application.  We expect that our results will generalise to the multi-layer case, with increasing numbers of layers exhibiting contrasting properties potentially providing  additional flexibility for targeting a specific release profile.  Finally, as we reduce the thickness of each layer, in the limit we can obtain a continuously changing porosity.  Whilst we acknowledge that this may provide even more flexibility in terms of controlling the release, the model we consider is a useful starting point to assess the effect of variable porosity.  Additionally, the manufacturing of such a system is likely to be very challenging given the typical coating thicknesses.

We would like to emphasise that we have made a number of simplifications in this work.  Perhaps the most significant is the assumption that drug is transported via a diffusive mechanism only. Depending on the particular coating material under consideration, it may be more appropriate to account for: polymer-drug interactions; diffusion through the solid phase; erosion;  swelling and/or degradation.  Additionally, in cases where fluid penetration into the coating is slow and/or the drug in question is poorly soluble, then the model may need to account for the drug dissolution process.  Nevertheless, the model we have provided here will be relevant in a number of drug delivery cases. Now that we have established that variable porosity coatings for drug-eluting devices are worth further consideration, we will seek to study some of these additional features in future work.

\section*{Acknowledgments}
Sean McGinty would like to acknowledge the funding provided by EPSRC under grant nos. EP/J007242/1 and EP/J007579/1. David King would like to acknowledge funding provided by EPSRC under grant number EP/M506539/1.




\end{document}